\documentclass[fleqn,usenatbib]{mnras}

\usepackage{newtxtext,newtxmath}

\usepackage[T1]{fontenc}
\usepackage{ae,aecompl}

\usepackage{graphicx}	
\usepackage{amsmath}	

\usepackage{amssymb}	
\usepackage[dvipsnames]{xcolor}     
\usepackage[]{hyperref} 
\usepackage{braket} 
\usepackage{comment} 
\usepackage{courier} 
\usepackage{subfig} 
\usepackage{siunitx} 
\usepackage{multirow} 

\newcommand{\rp}{\mbox{\textsc{rpicard}}}
\newcommand{\meqs}{\mbox{\textsc{meqsilhouette}}}
\newcommand{\meqsvtwo}{\mbox{\textsc{meqsv2}}}
\newcommand{\casa}{\mbox{\textsc{casa}}}
\newcommand{\aips}{\mbox{\textsc{aips}}}
\newcommand{\lpcal}{\mbox{\textsc{lpcal}}}
\newcommand{\poltools}{\mbox{\textsc{poltools}}}
\newcommand{\polsol}{\mbox{\textsc{polsolve}}}

\newcommand{\ehtim}{\mbox{\textsc{eht-imaging}}}
\newcommand{\sbrane}{\mbox{\textsc{scatterbrane}}}
\newcommand{\clean}{\mbox{\textsc{clean}}}
\newcommand{\atm}{\mbox{\textsc{atm}}}
\newcommand{\vlbimon}{\mbox{\textsc{vlbimonitor}}}
\newcommand{\symba}{\mbox{\textsc{symba}}}
\newcommand{\wsclean}{\mbox{\textsc{wsclean}}}
\newcommand{\codex}{\mbox{\textsc{codex-africanus}}}
\newcommand{\dask}{\mbox{\textsc{dask}}}
\newcommand{\daskms}{\mbox{\textsc{dask-ms}}}

\title[MeqSilhouette v2: Synthetic data generation for the EHT]{MeqSilhouette v2: Spectrally-resolved polarimetric synthetic data generation for the Event Horizon Telescope}

\author[Natarajan et al.]{Iniyan Natarajan$^{1,2,3}$\thanks{E-mail: iniyan.natarajan@wits.ac.za},
Roger Deane$^{1,4}$,
Iv\'an Mart\'i-Vidal$^{5,6}$,
Freek Roelofs$^{7,8,9}$,
\newauthor
Michael Janssen$^{10,9}$,
Maciek Wielgus$^{8,7,10}$,
Lindy Blackburn$^{8,7}$,
Tariq Blecher$^{3,2}$,
\newauthor
Simon Perkins$^{2}$,
Oleg Smirnov$^{3,2}$,
Jordy Davelaar$^{11,12,9}$,
Monika Moscibrodzka$^{9}$,
\newauthor
Andrew Chael$^{13}$, 
Katherine L. Bouman$^{14}$,
Jae-Young Kim$^{10,15}$,
Gianni Bernardi$^{16,3,2}$,
\newauthor
Ilse van Bemmel$^{17}$,
Heino Falcke$^{9}$,
Feryal {\"O}zel$^{18}$,
Dimitrios Psaltis$^{18}$
\\
$^{1}$Wits Centre for Astrophysics, School of Physics, University of the Witwatersrand, Private Bag 3, 2050, Johannesburg, South Africa\\
$^{2}$South African Radio Astronomy Observatory, Observatory 7925, Cape Town, South Africa\\
$^{3}$Centre for Radio Astronomy Techniques and Technologies, Department of Physics and Electronics, Rhodes University, Makhanda 6140,\\ South Africa\\
$^{4}$Department of Physics, University of Pretoria, Hatfield, Pretoria, 0028, South Africa\\
$^{5}$Departament d'Astronomia i Astrof\'{i}sica, Universitat de Val\`{e}ncia, C. Dr. Moliner 50, E-46100 Burjassot, Val\`{e}ncia, Spain\\
$^{6}$Observatori Astron\`{o}mic, Universitat de Val\`{e}ncia, C. Catedr\`{a}tico Jos\'{e} Beltr\'{a}n 2, E-46980 Paterna, Val\`{e}ncia, Spain\\
$^{7}$Center for Astrophysics | Harvard \& Smithsonian, 60 Garden Street, Cambridge, MA 02138, USA\\
$^{8}$Black Hole Initiative at Harvard University, 20 Garden Street, Cambridge, MA 02138, USA\\
$^{9}$Department of Astrophysics, Institute for Mathematics, Astrophysics and Particle Physics (IMAPP), Radboud University, PO Box 9010, \\6500 GL Nijmegen, The Netherlands\\
$^{10}$Max-Planck-Institut f\"ur Radioastronomie, Auf dem H\"ugel 69, D-53121 Bonn, Germany\\
$^{11}$Department of Astronomy and Columbia Astrophysics Laboratory, Columbia University, 550 W 120th Street, New York, NY 10027, USA\\
$^{12}$Center for Computational Astrophysics, Flatiron Institute, 162 Fifth Avenue, New York, NY 10010, USA\\
$^{13}$Princeton Center for Theoretical Science, Jadwin Hall, Princeton University, Princeton, NJ 08544, USA\\
$^{14}$California Institute of Technology, 1200 East California Boulevard, Pasadena, CA 91125, USA\\
$^{15}$Korea Astronomy and Space Science Institute, Daedeok-daero 776, Yuseong-gu, Daejeon 34055, Republic of Korea\\
$^{16}$INAF-Istituto di Radioastronomia, via Gobetti 101, 40129 Bologna, Italy\\
$^{17}$ Joint Institute for VLBI ERIC, Oude Hoogeveensedijk 4, 7991 PD Dwingeloo, Netherlands\\
$^{18}$ Department of Astronomy and Steward Observatory, University of Arizona, 933 North Cherry Avenue, Tucson, AZ 85721, USA\\
}

\date{Accepted XXX. Received YYY; in original form ZZZ}

\pubyear{2021}

\begin{document}
\label{firstpage}
\pagerange{\pageref{firstpage}--\pageref{lastpage}}
\maketitle

\begin{abstract}
We present MeqSilhouette v2.0 (MeqSv2), a fully polarimetric, time-and frequency-resolved synthetic data generation software for simulating millimetre (mm) wavelength very long baseline interferometry (VLBI) observations with heterogeneous arrays. Synthetic data are a critical component in understanding real observations, testing calibration and imaging algorithms, and predicting performance metrics of existing or proposed sites. MeqSv2 applies physics-based instrumental and atmospheric signal corruptions constrained by empirically-derived site and station parameters to the data. The new version is capable of applying instrumental polarization effects and various other spectrally-resolved effects using the Radio Interferometry Measurement Equation (RIME) formalism and produces synthetic data compatible with calibration pipelines designed to process real data. We demonstrate the various corruption capabilities of MeqSv2 using different arrays, with a focus on the effect of complex bandpass gains on closure quantities for the EHT at 230 GHz. We validate the frequency-dependent polarization leakage implementation by performing polarization self-calibration of synthetic EHT data using PolSolve. We also note the potential applications for cm-wavelength VLBI array analysis and design and future directions.
\end{abstract}



\section{Introduction}
\label{sec:intro}
Very long baseline interferometry (VLBI) enables the highest angular resolution in astronomy, on the order of milli-arcseconds (mas) to micro-arcseconds ($\mu$as), by operating radio antennas separated by thousands of kilometres synchronously using atomic clocks. The Event Horizon Telescope \citep[EHT,][]{eht2} is a global mm-VLBI array whose principal goal is to spatially resolve the supermassive black holes at the cores of the Milky Way galaxy (Sgr\,A$^*$) and M87, and image their \emph{shadows}, the depression in observed intensity inside the apparent boundary of the black hole \citep[e.g.][]{Falcke2000,dexter2012,psaltis2015,moscibrodzka2016}, together with a bright crescent-shaped emission ring \citep[e.g.][]{bromley2001,broderick2009,kamruddin2013,lu2014}. The EHT 2017 campaign has yielded total intensity images of the shadow of the black hole at the centre of M87 at 230 GHz \citep{eht1, eht4, eht6}. Assuming statistically preferred geometric crescent models and general-relativistic magnetohydrodynamic (GRMHD) models, measurements of physical properties such as the diameter of the shadow ($42\pm 3\, \upmu$as) and angular size of one gravitational radius ($3.8\pm 0.4\, \upmu$as) have been obtained \citep{eht4,eht5,eht6}. These measurements correspond to a mass of $6.5\pm0.7\times 10^9{\mathrm M}_{\odot}$ at the estimated distance $16.8^{+0.8}_{-0.7}$ Mpc \citep{eht6}, consistent with prior mass measurements based on stellar dynamics \citep{gebhardt2011}.

Synthetic data play a significant role in understanding the characteristics of an instrument, developing new algorithms for data analysis, and realistically representing the underlying physics that give rise to the observed data. 
Feasibility studies and identification of new sites for upgrading existing arrays such as the EHT, the Karl G Jansky Very Large Array (VLA) \citep{perley2011}, European VLBI Network (EVN) \citep{porcas2010}, East Asian VLBI Network (EAVN) \citep{antao2018}, the Atacama Large Millimeter Array (ALMA) \citep{matthews2018, goddi2019}, and MeerKAT \citep{jonas2009}, and for building new arrays such as the next-generation EHT (ngEHT) \citep{blackburnngeht2019}, the next-generation VLA (ngVLA) \citep{selina2018}, and the Square Kilometre Array (SKA) \citep[e.g.][]{schilizzi2007} can benefit greatly from realistic simulations of interferometric observations. These benefits include predictive analyses of new hardware enhancements, as well as testing and optimization of calibration and imaging algorithms and pipelines.
Powerful analyses can be performed by the average user when the synthetic data tools are user-friendly and can be seamlessly integrated with the calibration and analysis tools. \meqsvtwo{} has been designed with this as a guiding principle.

\citet{roelofs2020} provide a brief review of the various synthetic data generation approaches used for simulating EHT observations over the past decade. Most of these incorporate thermal noise arising from the characteristics of the instrument and the atmosphere as the only data corruption effect. More complex signal corruptions are introduced by \ehtim{} \citep{chael2016,Chael2018} and \meqs{} \citep[][and this paper]{tariq2017}, the two synthetic data generation packages used for generating synthetic M87 observations in \citet{eht1,eht4}. 

\ehtim{} can introduce randomly varying complex gains and elevation-dependent atmospheric opacity corruptions, and also simulates scattering due to the interstellar medium (ISM), which affects observations of Sgr\,A$^*$ \citep{johnson2016,johnson2018}.
Instrumental polarization is introduced using previous measurements of leakage characteristics of the EHT antennas \citep{johnsonetal2015} and the antenna gains are generated as random offsets sampled from a normal distribution with a standard deviation that is within a fixed percentage of the visibility amplitudes of actual EHT data.
\ehtim{} also adds randomized station-dependent inter-scan phases, that are kept coherent within a single scan to mimic the phases of fringe-fitted visibilities.

\meqs{} takes a complementary approach to synthetic data generation by introducing corruptions based on physical models and tuned to match empirical station measurements. 
It was designed to adapt the simulation and calibration techniques developed for metre and cm-wavelength observations to mm-VLBI observations. The version presented in \citet{tariq2017} could simulate simple Gaussian sources and narrow-field complex sky models and introduce physically-motivated tropospheric phase and amplitude corruptions, interstellar scattering using \sbrane{}\footnote{\url{https://github.com/krosenfeld/scatterbrane}.} \citep{johnson2015}, and time-variable antenna pointing errors. It has been significantly enhanced since then, first in step with the publication of the first results from 2017 EHT observations, and then with the development of \symba{} \citep{roelofs2020} and the first polarimetric results of the 2017 M87 observations \citep{eht7,eht8}. 

\meqs{}  \textsc{v2} (hereafter \meqsvtwo{}\footnote{``Measurement EQuation" (see section \ref{sec:rime}) + ``Silhouette" (referring to the black hole ``shadow").}) introduces full polarimetric, time-variable, spectrally-resolved synthetic data generation and corruption capabilities and is capable of handling wide-field source models with complex substructures. \meqsvtwo{} has been rewritten and expanded from \citet{tariq2017}. The code has been refactored to be fully compatible with the same pipelines used for the analysis of real EHT data. The pointing and atmospheric models have been rewritten to include more sophisticated effects. Source and instrumental polarization simulation capabilities have been introduced. \meqsvtwo{} also accounts for the effects of bandwidth on various propagation path effects at mm-wavelengths. These features facilitate a variety of studies for both the EHT and upcoming VLBI arrays, such as performing rotation measure (RM) synthesis studies \citep{brentjens2005} and multi-frequency synthesis imaging with the increasing fractional bandwith of the EHT, as well as the envisioned multi-band imaging at 230 GHz and 345 GHz for the ngEHT. The ability to vary instrumental polarization across the receiver bandwidth is crucial to take full advantage of ultra-wideband receivers and high dynamic-range polarimetric imaging. With the polarimetric primary beam module, full Stokes primary beam modelling for upcoming arrays such as the ngEHT can be undertaken. \citet{roelofs2020} seamlessly combines the functionality of \meqsvtwo{} and \rp{} \citep{janssen2019}, the \textsc{casa}-based VLBI pipeline for calibrating data from the EHT and other VLBI facilities. Synthetic data generated by both \ehtim{} and \symba{}, representing complementary approaches, have been found to be consistent with each other \citep{eht4}.

In this paper, we present the components of \meqsvtwo{} and illustrate its simulation capabilities. In particular, we illustrate the new polarimetric and spectral resolution capabilities using synthetic data and validate them. We study the effects of bandpass gains on closure quantities, which can limit or bias constraints on intrinsic source structure asymmetry, by generating synthetic data with realistic bandpasses. The polarimetric capabilities of \meqsvtwo{} are validated using \polsol{}, a \casa{} task developed for polarization calibration of VLBI observations \citep{polsolveRef}. Unlike conventional VLBI polarimetric calibration software packages such as \lpcal{} \citep{leppanen1995} that use a linear approximation to model polarization leakage, \polsol{} uses a non-linear model derived from the full RIME to handle high leakage values for specialised cases such as the EHT. In addition, \polsol{} uses a combined multi-source model when the parallactic angle coverage for individual calibrators is limited and can model the frequency-dependence of the leakage terms for calibrating large fractional bandwidths.

This paper is organized as follows: Section \ref{sec:rime} provides a brief overview of the RIME formalism that forms the basis of how \meqsvtwo{} models visibilities to make this a self-contained publication. Section \ref{sec:framework} provides a detailed account of the control flow of \meqsvtwo{}, its signal corruption capabilities, and their Jones matrix implementations. Section \ref{sec:polsolve} describes the \casa{} \polsol{} tool for estimating polarization leakage in heterogenous VLBI arrays.
Section \ref{sec:demo} demonstrates the synthetic data generation capabilities of \meqsvtwo{} for three different mm-wave telescopes. Section \ref{sec:bandpass} quantifies the effect of complex bandpass gains on EHT observations at 230 GHz and section \ref{sec:poldemo} demonstrates and validates the polarimetric simulation capabilities of \meqsvtwo{} using multiple polarized source models. Section \ref{sec:disc} provides a general discussion on the potential uses of \meqsvtwo{} and section \ref{sec:final} summarises the results and future outlook.


\section{The Radio Interferometer Measurement Equation}
\label{sec:rime}
\citet{hbs1996} originally developed the mathematical formalism for describing radio polarimetry using the \emph{Jones} \citep{jones1941} and \emph{Mueller} \citep{mueller1948} calculi from optics. \citet{oms2011} extended this formulation to incorporate direction-dependent effects (DDEs) in calibration. Here we provide only a brief summary of the relevant aspects of this formalism and, given the range of notations in use, establish the notation used in this paper.

An interferometer produces four pairwise correlations between the voltage vectors from two stations $p$ and $q$ (each with two feeds $x$ and $y$), that can be arranged into the 2$\times$2 \emph{visibility matrix}\footnote{The factor of 2 is introduced in this equation to ensure that the brightness matrix (introduced shortly) becomes $1$ for a $1$ Jy unpolarized source \citep{oms2011}.}:
\begin{equation}
\label{eq:rime3}
\mathbfss{V}_{pq} = 2 \begin{pmatrix} \braket{v_{px}v^*_{qx}} && \braket{v_{px}v^*_{qy}} \\  
\braket{v_{py}v^*_{qx}} && \braket{v_{py}v^*_{qy}}
\end{pmatrix},
\end{equation}
where the angled brackets denote averaging over some small time and frequency bin, based on considerations of \emph{smearing} and \emph{decoherence}, field of interest, and processing requirements. In terms of the voltage two-vectors $\boldsymbol{v}_p$, equation (\ref{eq:rime3}) can be represented as
\begin{equation}
\begin{aligned}
\label{eq:rime4}
\mathbfss{V}_{pq} &= 2\; \Bigg \langle \begin{pmatrix} v_{px} \\ v_{py} \end{pmatrix} \begin{pmatrix} v_{qx}^*, v_{qy}^* \end{pmatrix} \Bigg \rangle
= 2 \braket{\boldsymbol{v}_p \boldsymbol{v}^H_q}, \\
&=2 \braket{\, \boldsymbol{J}_p\boldsymbol{e} (\, \boldsymbol{J}_q\boldsymbol{e})^H} = 2 \braket{\, \boldsymbol{J}_p (\boldsymbol{e}\boldsymbol{e}^H) \boldsymbol{J}_q^H},
\end{aligned}
\end{equation}
where $\boldsymbol{e}$ is the incoming electromagnetic wave, $\boldsymbol{J}_p$ are the 2$\times$2 \emph{Jones} matrices that describe any linear transformation acting on the incoming wave, and $H$ is the Hermitian conjugate. The matrix product $\boldsymbol{e}\boldsymbol{e}^H$ in equation (\ref{eq:rime4}) is related to the four Stokes parameters $I$, $Q$, $U$, and $V$ that describe the polarization state of electromagnetic radiation \citep{hbs1996,TMS2017} by the following relation\footnote{This equation is valid for linear (XY) feeds. See section \ref{subsec:skymodels} for circular (RL) feeds.}:
\begin{equation}
\label{eq:rime6}
2\, \begin{pmatrix} \braket{e_x e^*_x} && \braket{e_x e^*_y} \\  
\braket{e_y e^*_x} && \braket{e_y e^*_y}
\end{pmatrix}
= \begin{pmatrix} I+Q && U+iV \\ U-iV && I-Q \end{pmatrix} \equiv \mathbfss{B}.
\end{equation}
$\mathbfss{B}$ is the \textit{brightness matrix} which describes the intrinsic source brightness. $e_x$ and $e_y$ are the orthogonal polarizations as measured by the two feeds $x$ and $y$.

In the ideal case of corruption-free reception, the phase delay associated with signal propagation, denoted by the scalar \emph{K-Jones} matrix, is always present, giving rise to the \emph{source coherency}, $\mathbfss{X}_{pq}$:
\begin{equation}
\begin{aligned}
\label{eq:rime7}
\mathbfss{X}_{pq} &= K_p \mathbfss{B} K_q^H\,, \\
\mathrm{where}\, K_p &= e^{-2\pi i\phi_p} \begin{pmatrix}
1 && 0 \\ 0 && 1
\end{pmatrix}
\end{aligned}
\end{equation}
in which $\phi_p$ denotes the phase delay between the antenna $p$ and the reference antenna. In the presence of multiple discrete sources in the sky, taking into account the direction-dependence of the source coherency and some Jones matrices, the RIME generalizes to
\begin{equation}
\label{eq:rime8}
\mathbfss{V}_{pq} = G_p \left( \sum_{s} E_{sp}\, \mathbfss{X}_{spq}\, E_{sq}^H \right) G_q^H,
\end{equation}
where the summation is carried out over all the sources and $\boldsymbol{E}_{sp}$ and $\boldsymbol{G}_p$ denote generic direction-dependent effects (DDEs) and direction-independent effects (DIEs) respectively. \meqsvtwo{} does not simulate any DDEs for EHT observations, since, aside from scattering, there are no major DDEs that occur along the signal path.

\section{The MeqSilhouette Framework}
\label{sec:framework}
\begin{figure*}
\centering
 \includegraphics[scale=0.415]{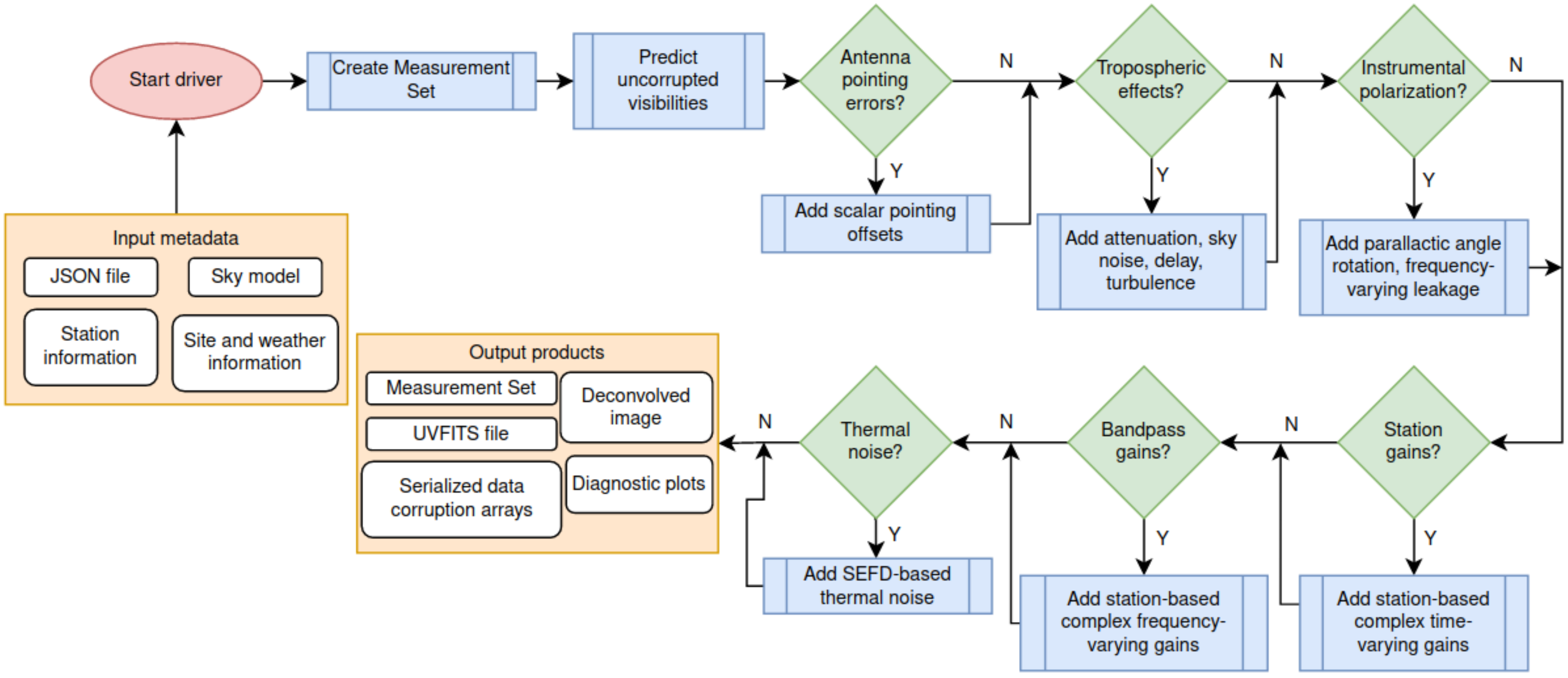}
 \caption[meqsv2-components]{Flowchart showing the basic components of synthetic data generation with \meqsvtwo{}. The inputs and outputs are shaded orange, the processes are shaded blue, and the decision boxes (diamonds) are shaded green. Multiple input configuration files are used and various output data products are produced. The input values are determined from empirical values obtained from the individual observations themselves, as well as \vlbimon{}. Each component in the diagram is explained in section \ref{sec:framework}.
}
\label{fig:meqsv2components}
\end{figure*}

\meqsvtwo{} was designed to use the Measurement Set\footnote{\url{https://casa.nrao.edu/Memos/229.html}.} (MS), a database format designed to store radio astronomical data in next-generation facilities such as JVLA, ALMA, MeerKAT, and SKA.
Fig. \ref{fig:meqsv2components} shows the basic layout and the components of a typical \meqsvtwo{} run.
Scattering by the ISM is not applied to the input sky models and is assumed to have been applied externally, simplifying the user interface compared to v1. \meqsvtwo{} uses a \emph{driver} script to set up the sequence of steps to be executed to generate the synthetic data. The \emph{framework} module contains the various functions necessary to create synthetic data, corrupt them, and optionally generate additional data products. The inputs to the driver script are presented as attribute-value pairs in a file in the \textsc{json} format \citep{crockford2006json} containing information on the source, weather, and antenna parameters necessary for computing various components of the RIME.
The Jones matrices are applied to the uncorrupted visibilities in the order in which they occur along the signal path \citep{meqtrees2010}, unless they are scalar, in which case they can be applied anywhere along the signal chain.
 
Advanced users may write their own driver scripts to tailor the basic strategy provided by \meqsvtwo{} for their own needs. For instance, in \symba{} \citep {roelofs2020}, we use this framework to create synthetic data that follow real EHT observing schedules using input VEX\footnote{\url{https://vlbi.org/vlbi-standards/vex}.} files (the scheduling protocol for VLBI experiments) and extend the pointing offset module to compute short and long-term pointing offsets mimicking the behaviour of EHT stations. \symba{} also performs additional a priori calibration so that the output more closely resembles the uncalibrated EHT data. \meqsvtwo{} can just as easily be used for simulating observations with other VLBI arrays (see section \ref{sec:demo}), although other propagation path effects that become significant at longer wavelengths may need to be implemented.

The following subsections explain the various modules of \meqsvtwo{}. The plots shown are obtained using a hypothetical 12-hour observing run using the EHT2017 array listed in Table \ref{tab:sec3stationlist} at an observing frequency of 230 GHz towards M87 ($\alpha_{\rm J2000} =  12^\mathrm{h}30^\mathrm{m}49^\mathrm{s}.42,\, \delta_{\rm J2000} = 12^{\circ}23'28''.04$).
\begin{table}
\caption{Physical sizes and mount specifications of EHT2017 stations participating in the simulations.}
\begin{center}
\begin{tabular}{lcc}
\hline
Station & Effective Diameter (m) & Mount type\\
\hline
ALMA$^\dagger$ & 70 & Alt-az\\
APEX & 12 & Alt-az+Nasmyth-Right\\
LMT & 32 & Alt-az+Nasmyth-Left\\
PV & 30 & Alt-az+Nasmyth-Left\\
SMT & 10 & Alt-az+Nasmyth-Right\\
JCMT & 15 & Alt-az\\
SMA$^\dagger$ & 16 & Alt-az+Nasmyth-Left\\
\hline
\end{tabular}
\end{center}
\label{tab:sec3stationlist}
$^\dagger$Single-antenna equivalent of phased arrays.
\end{table}
The SPT station from EHT2017 has been excluded since M87 is always below the horizon from the south pole.

\subsection{Input sky models}
\label{subsec:skymodels}
\meqsvtwo{} introduces the capability to generate synthetic visibilities from wide-field non-parametric images and retains the ability to input simple parametric source models in the form of ASCII text files describing point or Gaussian source models. Since images are gridded representations of the sky, we use \wsclean{}, a fast generic widefield imager \citep{wsclean2014}, to Fourier-invert the model image to the \emph{uv}-plane. Each polarized, time and frequency variable image frame is Fourier-inverted into the appropriate subset of the generated MS\footnote{The naming conventions for the image frames are explained in the official documentation.}. Parametric sky models are handled by \textsc{meqtrees} \citep{meqtrees2010}, which performs a direct Fourier transform of the sky model into the MS.

\begin{figure}
\centering
 \includegraphics[scale=0.5]{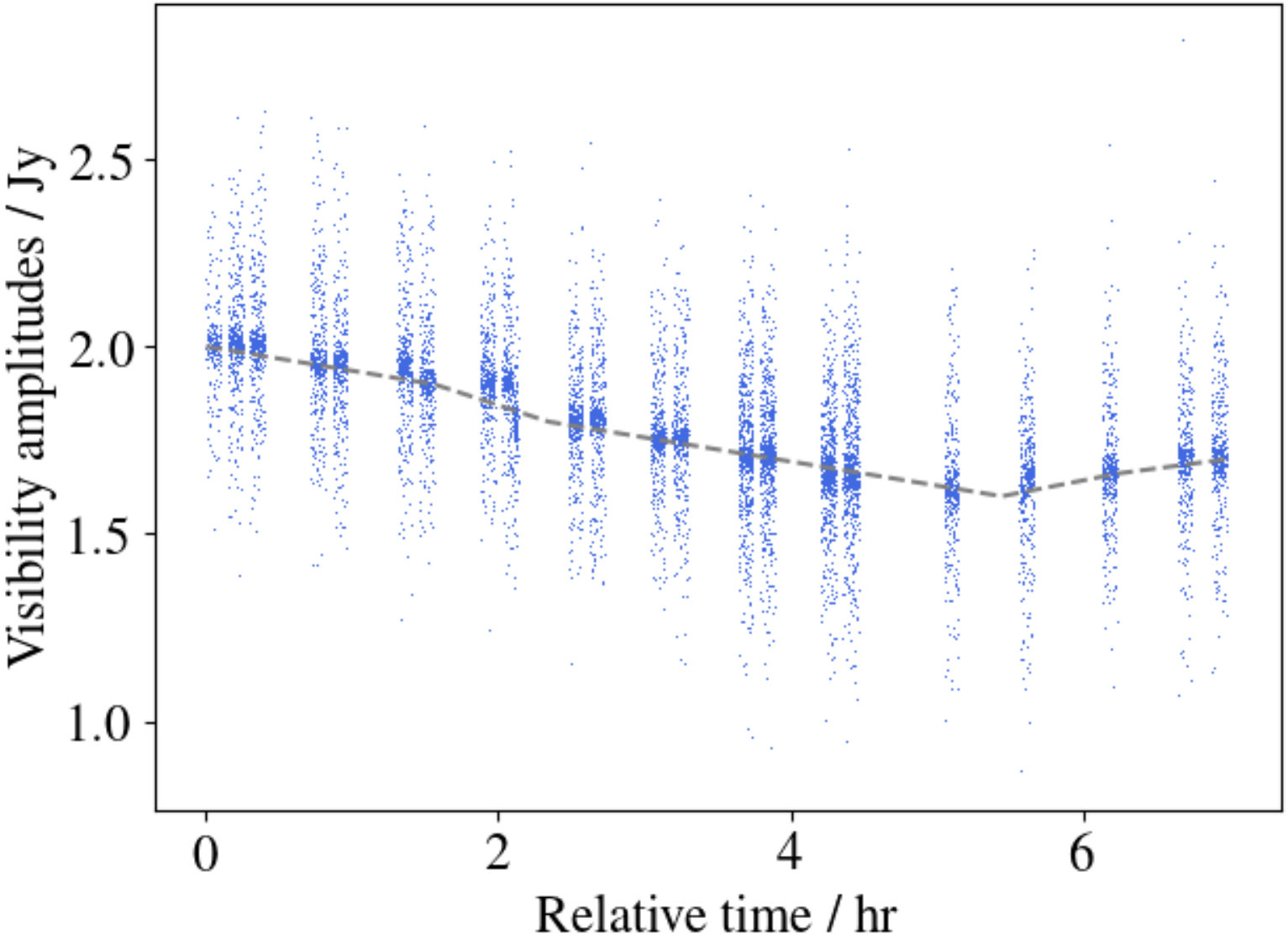}
 \caption[amp-vs-time]{Stokes $I$ visibility amplitudes for a point source with intrinsic time variability, as a function of time for one frequency channel.}
\label{fig:amp-vs-time}
\end{figure}

For the EHT, the visibilities are always computed in the circular polarization basis (RR, RL, LR, and LL) except in the case of ALMA which records signals in linear basis. In \meqsvtwo{}, we assume that the ALMA visibilities have been perfectly converted to circular polarization \citep{polconvert2015} so that the basis is uniform across all stations. Then, equation (\ref{eq:rime6}) takes the form
\begin{equation}
    \label{eq:circrime}
\mathbfss{B}_{\odot}  = 2\, \begin{pmatrix} \braket{e_r e^*_r} && \braket{e_r e^*_l} \\  
\braket{e_l e^*_r} && \braket{e_l e^*_l}
\end{pmatrix}
= \begin{pmatrix} I+V && Q+iU \\ Q-iU && I-V \end{pmatrix}.
\end{equation}
where `$\odot$' indicates circular polarization \citep{oms2011}.
Figure \ref{fig:amp-vs-time} shows the Stokes I visibility amplitudes for all baselines for a single channel of the data set generated using a point source sky model with intrinsic time variability. The scatter of the visibilities is due to thermal noise. 
 The capability to simulate time-varying sources is particularly useful for simulating sources exhibiting variability on timescales of minutes to hours, such as the radio source associated with the supermassive black hole Sagittarius\,A$^*$ (or Sgr\,A$^*$) located at the Galactic centre \citep[e.g.][]{lu2016}. In addition, studies on decoupling time-varying instrumental effects from source evolution could be undertaken.
 
\begin{figure}
\centering
 \includegraphics[scale=0.5]{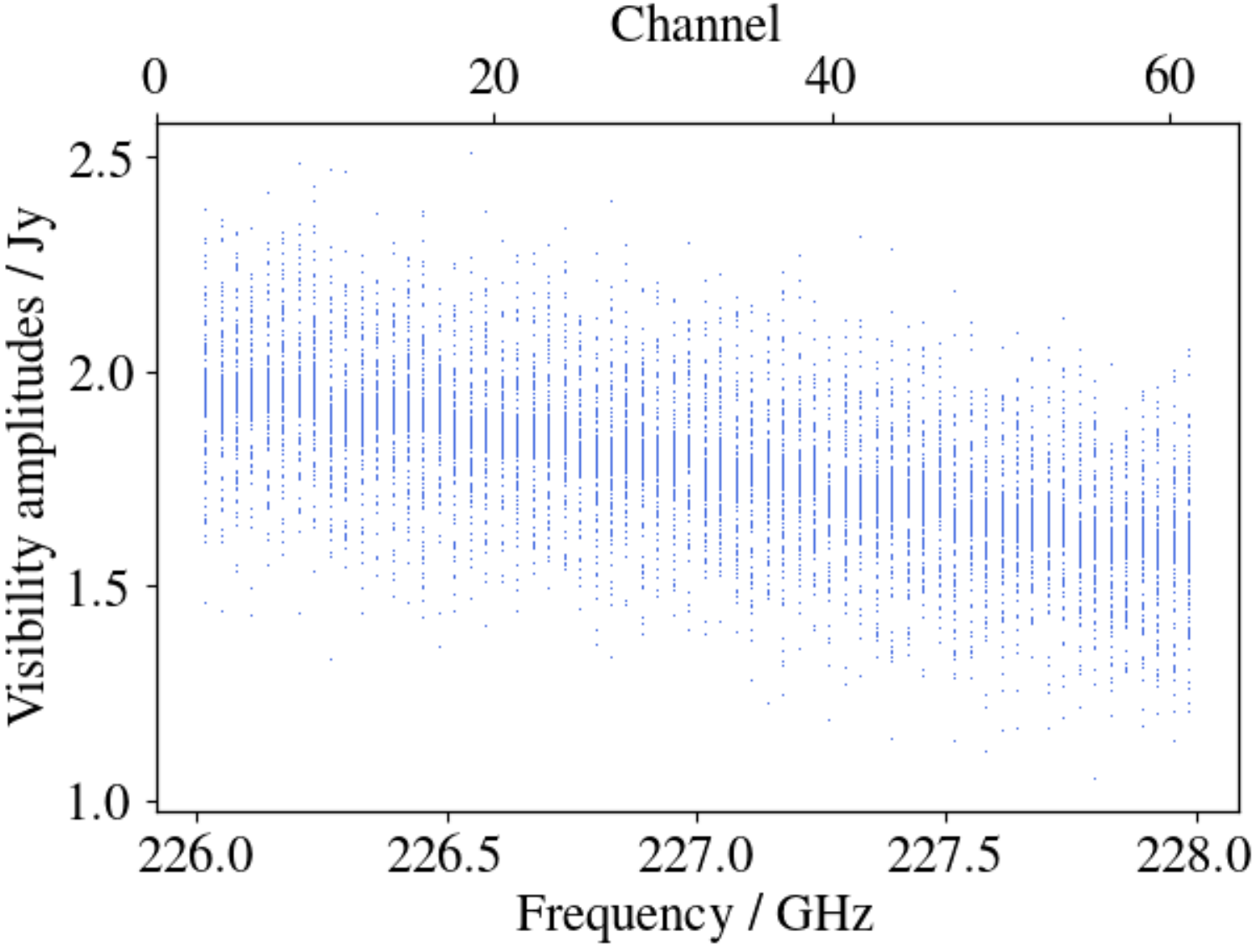}
 \caption[amp-vs-time]{Stokes $I$ visibility amplitudes of a point source with intrinsic frequency variability, as a function of frequency for 5 minutes of observation.}
\label{fig:amp-vs-freq}
\end{figure}

 Figure \ref{fig:amp-vs-freq} shows the Stokes $I$ visibilities of a point source with a steep spectral index, across a 2 GHz bandwidth divided into 64 channels, centred at 227 GHz.
The data shown correspond to a 5-minute subset of the entire observation for all channels. As before, the scatter seen is due to thermal noise.

\subsection{Antenna pointing errors}
\label{subsec:ptgerr}
Several factors cause antennas to mispoint and modify its gain response, such as differential flexure of the antenna, wind pressure, and thermal expansion. Moreover, errors in drive mechanics or the telescope control software also introduce pointing errors, which attenuate the measured visibility amplitudes, $|V_{pq}|$. Even small offsets in antenna pointing can cause significant reduction in visibility amplitudes at mm-wavelengths.

Formulated in terms of the RIME, this effect is represented by a time and direction-dependent antenna-based E-Jones matrix \citep{oms2011a}:
\begin{equation}
\label{eq:pointing}
E_p(l, m) = E(l+\delta l_p, m+\delta m_p),
\end{equation}
where $\delta l$ and $\delta m$ are the time-variable offsets in the $lm$-plane. \meqsvtwo{} provides a WSRT-derived analytic $\cos^3$ beam (from the previous version) and a Gaussian primary beam profile. The Gaussian profile has the advantage that it can be conveniently described by a single parameter, the full width at half maximum (FWHM), and does not deviate from the more complex Bessel function from the centre up to very close to the first null \citep{middelberg2013}. This assumption is justified in the context of EHT, since, in most cases, the field of interest is much smaller than the location of the first null.

The offsets per pointing epoch (or scan) for station $p$, $\rho_p$, are drawn from the normal distribution $\mathcal{N}(0\, , \mathcal{P}_{{\rm rms}, p})$. $\mathcal{P}_{\rm rms}$ values depend on station and weather characteristics and are determined based on empirical measurements. The full width at half maximum (FWHM) of the primary beam $\mathcal{P}_{{\rm FWHM}, p}$ of each antenna at 230 GHz is scaled to the centre frequency of the observation. The beam model with the pointing errors is given by
\begin{equation}
\begin{aligned}
\label{eq:ptggauss}
    E_p &= \exp \Bigg( -\frac{1}{2} \Bigg[\frac{\rho_p}{(\mathcal{P}_{{\rm FWHM}, p}/2\sqrt{2\ln 2})}\Bigg]^2 \Bigg)\, , \\
    \mathrm{where}\, \rho_p &= \sqrt{\delta l_p^2 + \delta m_p^2}.
\end{aligned}
\end{equation}
The $2\sqrt{2\ln 2}$ factor arises due to the relationship $\mathrm{FWHM} = 2\sqrt{2\ln 2}\sigma$, where $\sigma$ is the standard deviation. Updating equation (\ref{eq:ptggauss}) is all that is required to implement more complex, asymmetric primary beam patterns. The corresponding E-Jones matrix is given by 
\begin{equation}
    E_p = \begin{pmatrix} E_{pa}(l, m) && 0 \\ 0 && E_{pb}(l, m) \end{pmatrix}.
\end{equation}
This is implemented as a scalar term ($E_{pa} = E_{pb}$) that can commute with other components of the RIME.

\begin{figure}
 \includegraphics[scale=0.36875]{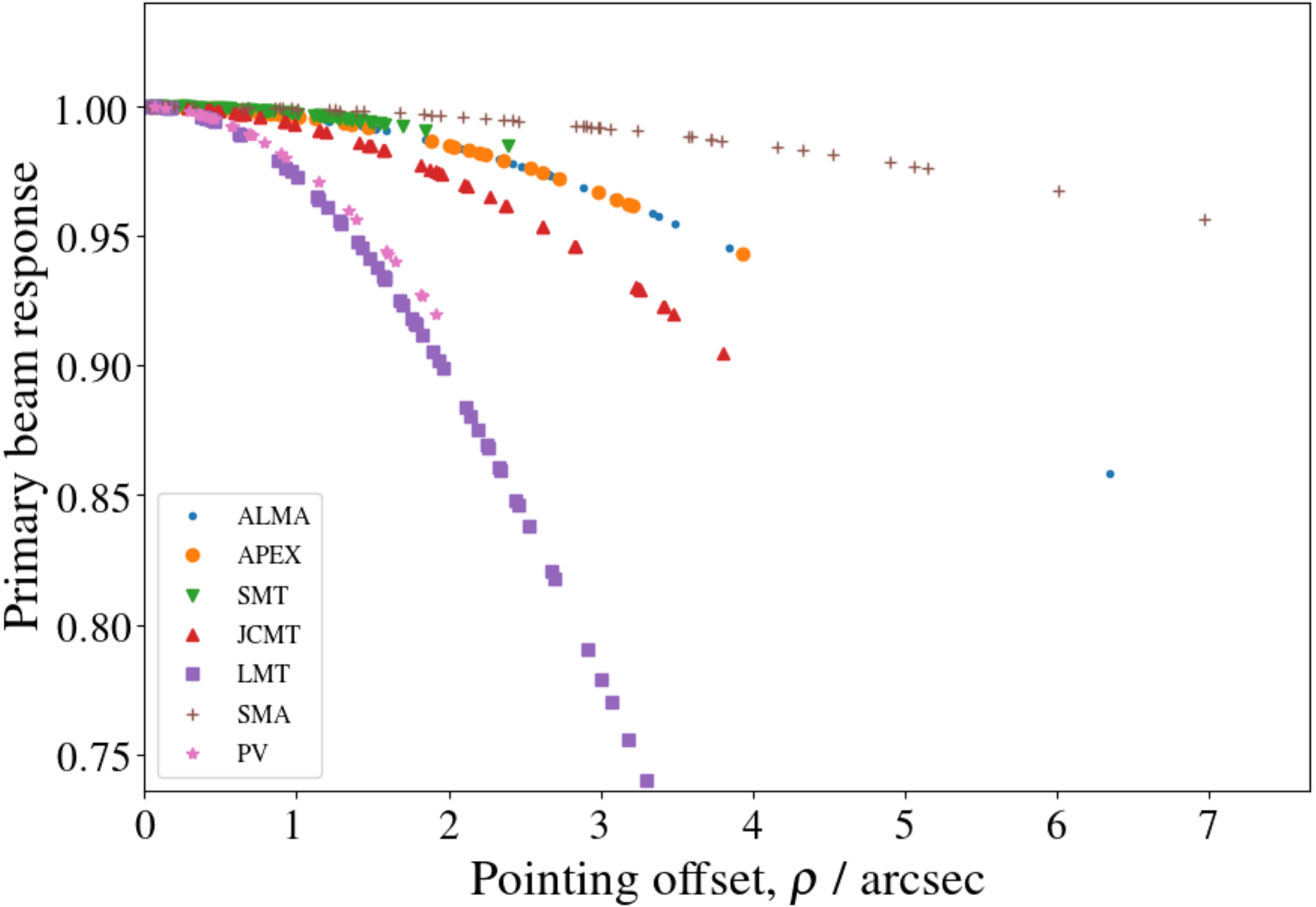}
 \caption[meqsv2-components]{The true primary beam response pattern for the EHT stations at 230 GHz, as a function of the pointing offset, $\rho$. Each point corresponds to one realisation of the pointing offset.}
\label{fig:ptg1}
\end{figure}

Figure \ref{fig:ptg1} shows an example simulation of pointing offsets, $\rho_p$, and how they affect the primary beam response of the EHT2017 antennas.
SMA is the least affected owing to its large beam size, while LMT is the most affected due its narrow primary beam.

\subsection{Tropospheric effects}
\label{subsec:trop}

The troposphere has a significant effect on signal propagation at mm wavelengths \citep{carilli1999}. \meqsvtwo{} models three significant tropospheric effects -- (i) the signal attenuation due to atmospheric opacity $\tau$, (ii) the increased system temperature due to atmospheric emission at the sky brightness temperature $T_{\mathrm{sky}}$, and (iii) the phase fluctuations due to atmospheric turbulence. These effects are separated into mean and turbulent components, with the mean component further divided into wet (due to water vapour) and dry components. A detailed treatment of the propagation fundamentals can be found in \citet{tariq2017}. Here we provide a summary along with the updated equations and algorithms implemented in \meqsvtwo{}.

\subsubsection{Mean troposphere}
\label{sss:meantrop}
At mm-wavelengths, the mean component of the troposphere causes signal attenuation due to absorption and time delays, along with a pseudo-continuum opacity that increases with frequency. The H$_2$O and O$_2$ lines cause significant absorption above 100 GHz, peaking in the THz regime \citep{TMS2017}. \meqsvtwo{} retains the use of the Atmospheric Transmission at Microwaves \citep[\atm{},][]{pardo2001} software from \citet{tariq2017} to compute the mean opacity, sky noise, and the mean component of the delays.
\atm{} implements radiative transfer through a static atmosphere. In the absence of scattering, the radiative transfer equation for unpolarized radiation is given by
\begin{equation}
    \label{eq:radtrans}
    \frac{\mathrm{d}I_{\nu}(s)}{\mathrm{d}s} = \epsilon_{\nu}(s) - \kappa_{\nu}(s)I_{\nu}(s)\,,    
\end{equation}
where $s$ is the coordinate along the direction of the signal path through the atmosphere, $I_{\nu}(s)$ is the specific intensity at frequency $\nu$, and $\epsilon_{\nu}$ and $\kappa_{\nu}$ are respectively the emission and absorption coefficients of the atmosphere. The absorption coefficient $\kappa_{\nu}$ is calculated as a function of frequency and altitude, enabling equation (\ref{eq:radtrans}) to be integrated along the signal path to obtain mean opacity and sky brightness temperature.

\begin{figure}
 \includegraphics[scale=0.55]{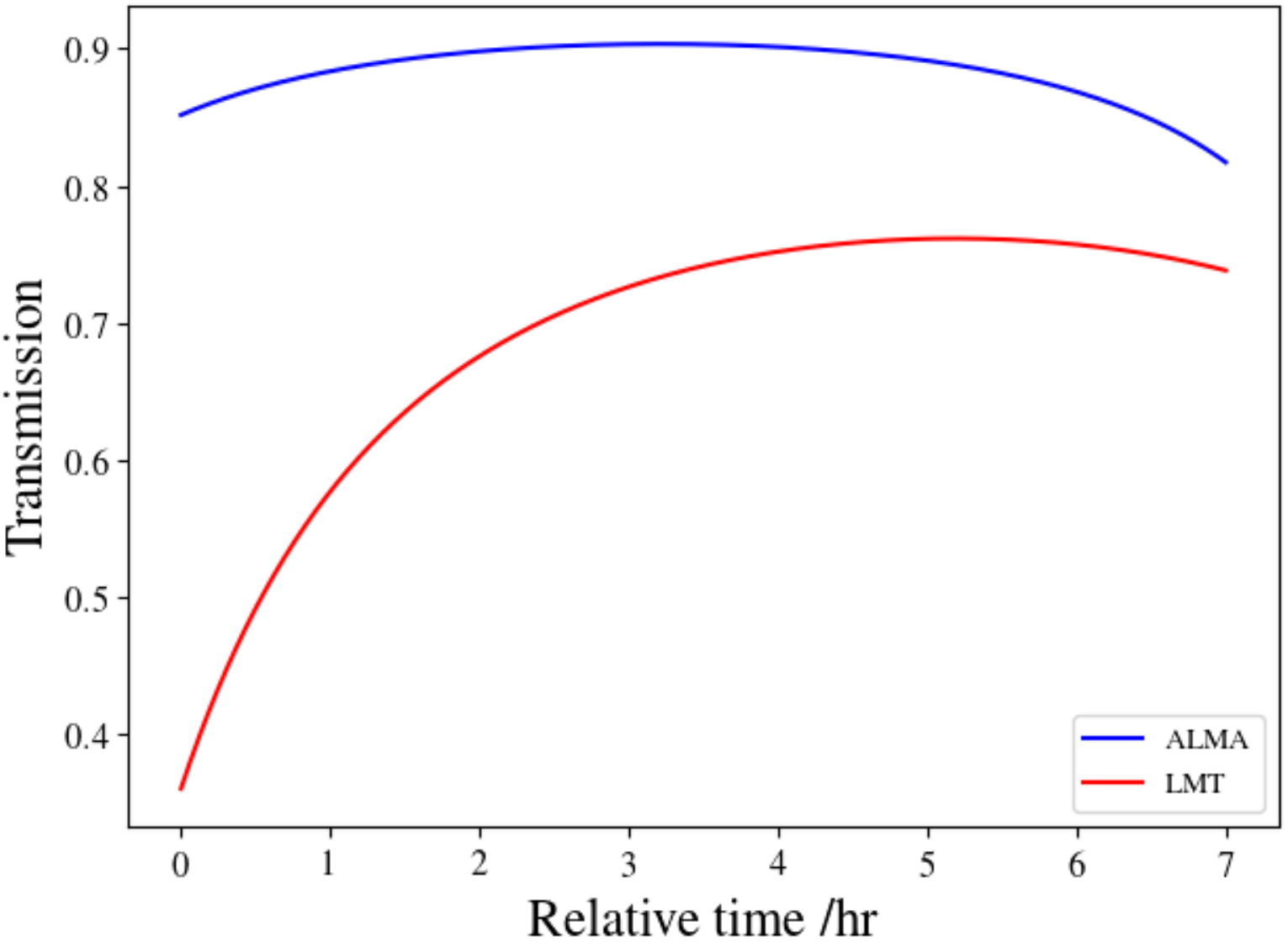}
 \caption{Variation in transmission with time at the sites of ALMA and LMT, during a 7-hour tracking of M87. }
\label{fig:elevtrans}
\end{figure}

An example of the elevation-dependent transmission for ALMA and LMT over a 7-hour observing track towards the direction of M87 is shown in Figure \ref{fig:elevtrans}.
The mean time delay is obtained from the absorption coefficient using the Kramers-Kr\"onig relations \citep[e.g.][]{kronig1926}, for which the necessary atmospheric temperature and pressure profiles are calculated based on site-dependent precipitable water vapour (PWV) depth and ground temperature and pressure \citep{pardo2001}.

The a priori correlator model used by the EHT corrects for various effects including the elevation-dependent variations in the intra-scan delays \citep{eht2}. Optionally, \meqsvtwo{} simulates this correction by introducing a constant residual delay per scan, that is equal to the mean delay averaged over all frequency channels.

Integrating equation (\ref{eq:radtrans}) also yields sky brightness temperature, which contributes to an increase in the system temperature. The resultant increase in the sky noise is implemented as an opacity and elevation-dependent noise component. This noise, along with the thermal noise component, is accounted for in the total noise budget used for visibility weighting in \meqsvtwo{} (see section \ref{subsec:thermalnoise}).

\subsubsection{Turbulent troposphere}
\label{sss:turbtrop}
Turbulence in the  troposphere introduces phase instabilities in the measured visibilities. The spatial distribution of water vapour in the troposphere evolves rapidly, reducing the coherence timescale to $\sim\negthickspace10$ s \citep{TMS2017}. This limits the coherent averaging time (and hence the S/N) of uncalibrated visibilities, and renders conventional calibration procedures (with interleaved observations of calibrator and target source) ineffective. Self-calibration is also limited by the S/N required for fringe-fitting.

\meqsvtwo{} models these phase variations, $\delta \phi(t, \nu)$, by assuming a thin, Kolmogorov-turbulent phase screen moving with constant velocity \citep[e.g.][]{johnson2015}. This method is applicable to any situation in which the troposphere induces delays in electromagnetic wave propagation \citep[e.g.][]{treuhaft1987}. The phase offsets introduced can be described by the phase structure function given by \citep{carilli1999}
\begin{equation}
\label{eq:turb1}
\mathcal{D}_{\phi}(\boldsymbol{x}, \boldsymbol{x'}) = \langle[\phi(\boldsymbol{x}+\boldsymbol{x'})-\phi(\boldsymbol{x})]^2\rangle\,,
\end{equation}
where $\boldsymbol{x}$ and $\boldsymbol{x'}$ are points on the screen and the angled brackets denote ensemble average. This equation can be reasonably approximated by a power law \citep[e.g.][]{armstrong1995}:
\begin{equation}
\label{eq:turb2}
\mathcal{D}_{\phi}(r) \approx \bigg( \frac{r}{r_0} \bigg)^\beta\,,
\end{equation}
where $r^2=(\boldsymbol{x}-\boldsymbol{x'})^2$ and $r_0$ is the phase coherence scale at which $\mathcal{D}_{\phi}(r_0) = 1\ \mathrm{rad}$.

Scattering can be classified as \textit{strong} or \textit{weak} based on the relationship between $r_0$ and the Fresnel scale, $r_F$, defined as $\sqrt{\lambda D_{\mathrm{os}}/2\pi}$, where $D_{\mathrm{os}}$ is the distance between the observer and the scattering screen. The radiative power measured at a given point originates from a single region of area $A_{\mathrm{weak}} = \pi r_F^2$ when $r_F\ll r_0$ and from multiple disconnected zones each of area $A_{\mathrm{strong}} = \pi r_0^2$ when $r_F \gg r_0$ \citep{narayan1992}. Empirical estimates of $r_0$ fall in the range of $\sim 50-500$ m above Mauna Kea \citep{Masson1994} and $\sim 90-700$ m above Chajnantor \citep{Radford1998}. At $\lambda=1.3$ mm and $D_{\mathrm{os}}=2$ km (the water vapour scale height), $r_F \approx 0.64$ m ($\ll r_0$), and hence scattering falls under the weak regime.

Equation (\ref{eq:turb2}) may be rewritten with explicit time-dependence as $D_{\phi}(t)=D_{\phi}(r)|_{r=\boldsymbol{v}t}$, where $\boldsymbol{v}$ is the bulk transverse velocity of the phase screen \citep{coulman1985}. Assuming that the coherence timescale when observing towards the zenith, $t_{\mathrm c} = r_0/|\boldsymbol{v}|$ \citep{treuhaft1987,tariq2017}, we get
\begin{equation}
\label{eq:turb3}
\mathcal{D}_{\phi}(t) \approx \bigg( \frac{t}{t_c} \bigg)^\beta\,.
\end{equation}
Both Kolmogorov theory and empirical measurements show that the exponent $\beta$ should be equal to $5/3$ when $r<\Delta h$, where $\Delta h$ is the thickness of the turbulent layer which is $\sim 1$ km \citep{carilli1997}. The processed Field-of-View (FoV) of the EHT is about $100 \times 100\;\mu$as$^2$ and is much smaller than $r_0$, allowing us to represent it as a diagonal Jones matrix in the RIME.

The antenna-based turbulent phase errors manifest as a series of correlated, normally distributed random variables in time. This time-series, \{$\delta \phi^\prime(t)$\}, can be constructed as follows \citep[e.g.][]{rasmussen2006}. From the structure function $\mathcal{D}$ we construct the covariance matrix $\Sigma$. Since $\Sigma$ is symmetric and positive definite, the lower triangular matrix, $L$, resulting from the Cholesky decomposition of $\Sigma$ (where $\Sigma = LL^T$) can be applied to a time-series of the desired length with zero mean and unit variance to arrive at correlated random samples.

We assume a simple linear scaling with frequency across the bandwidth since the wet dispersive path delay is not more than a few per cent of the non-dispersive component at mm-wavelengths \citep{curtis2009}. Also taking into account the airmass towards the horizon when observing away from the zenith, the phase error time-series for antenna $p$ becomes
\begin{equation}
    \label{eq:turb4}
    \{{\delta \phi_p(t, \nu)}\} = \frac{1}{\sqrt{\sin({\theta_{\mathrm{el}}(t)})}} \{\delta\phi^{\prime}_p(t)\} \big(\frac{\nu}{\nu_0}\big)\,,
\end{equation}
where $\nu$ is the list of channel frequencies, $\nu_0$ is the reference frequency, taken to be the lowest channel frequency, and $\theta_{\mathrm{el}}$ is the elevation angle.

Since VLBI stations are typically located hundreds or thousands of kilometres from each other, the tropospheric corruptions over individual stations are uncorrelated with each other and must be simulated independently. This is not strictly true for the short-baseline pairs of ALMA-APEX (2.6 km) and JCMT-SMA (160 m) in the EHT, for which the turbulence may be correlated since the baseline lengths are so short as to be comparable to $r_0$ \citep[e.g.][]{carilli1997}. Currently, intra-site baseline correlations are not simulated and the phase errors are generated independently for these stations.

\subsection{Instrumental polarization}
\label{subsec:panddjones}
The two polarization feeds nominally measure two orthogonal polarization states of the incoming wave, in either circular (RL) or linear (XY) bases. In practice, mechanical imperfections in the feed or electronic imperfections in the signal path cause signals from each independent signal path to leak into the other \citep{hbs1996,hbs1996a}. Additionally, the possible rotation of the feeds by the parallactic angle depending on the antenna mount type must be taken into account.

\subsubsection{Parallactic angle rotation}
\label{subsubsec:parang}
\begin{figure}
 \includegraphics[scale=0.36875]{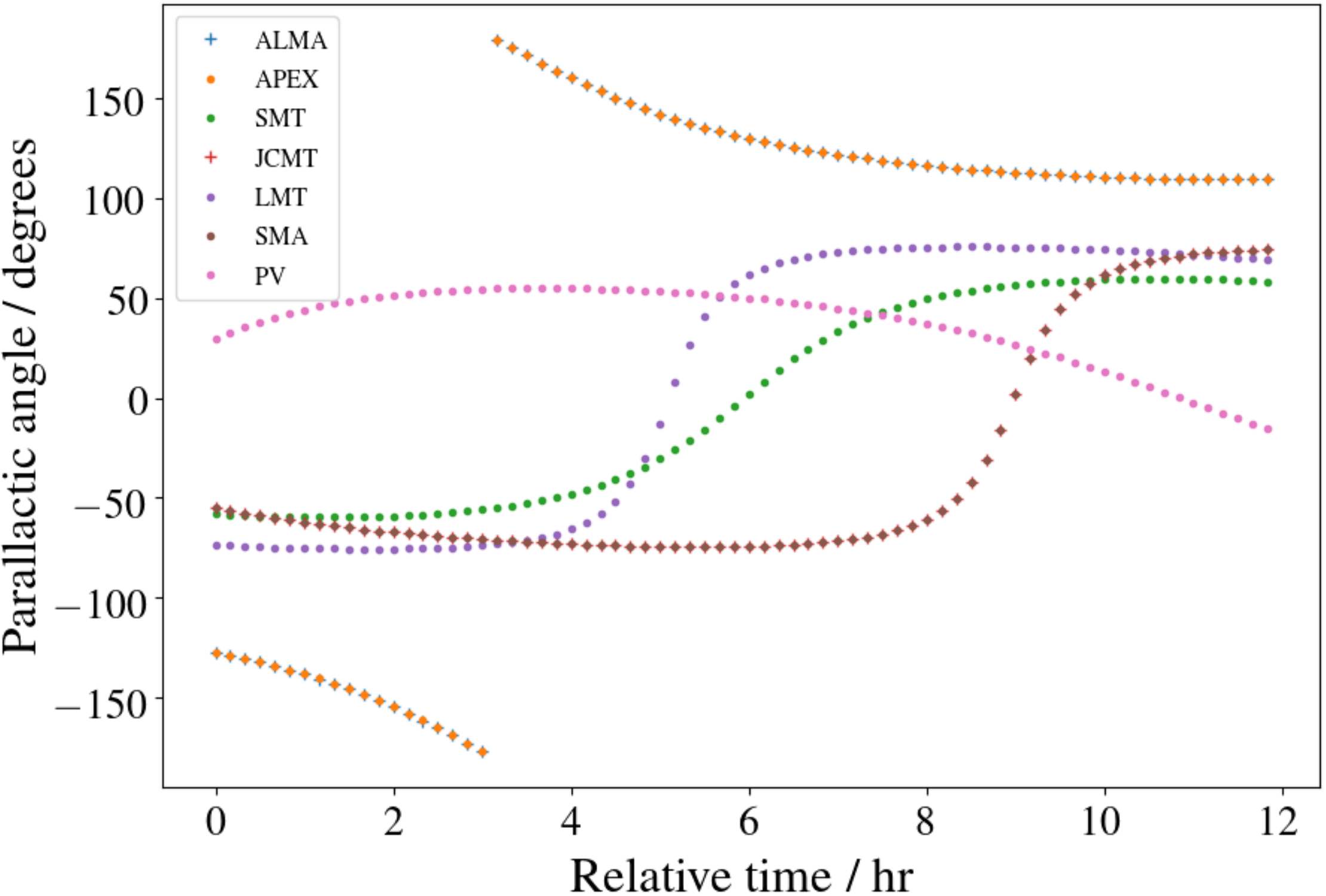}
 \caption{Evolution of the parallactic angles of EHT 2017 array stations over 12-hours in the direction of M87.}
\label{fig:parangplot}
\end{figure}

The mount type of the antenna determines the rotation of the feeds as seen from the sky \citep[e.g.][]{dodson2009}. For equatorially mounted antennas, the orientation of the primary beam with respect to the sky remains static. For alt-az mounted antennas, the beam rotates as the source is tracked over the sky, and the feeds rotate by the \emph{parallactic angle} ($\chi_{_{pa}}$) given by
\begin{equation}
    \chi_{_{pa}} = \arctan \bigg( \frac{\sin(\Theta)\cos(l)}{\cos(\delta)\sin(l)-\cos(\Theta)\cos(l)\sin(\delta)} \bigg)\, ,
\end{equation}
where $\Theta$ is the hour angle, $\delta$ is the Declination, and $l$ is the latitude. For alt-az mounted antennas with Nasmyth focus, the more generic \emph{feed angle} ($\chi_{_p}$) is used in place of parallactic angle, which incorporates the elevation of the antenna. Depending on whether the tertiary mirror focuses the light to the right or to the left, the elevation angle ($\chi_{_{el}}$) is added or subtracted respectively,
\begin{equation}
\label{eq:feedangle}
\begin{aligned}
    \chi_{_p} &= \chi_{_{pa}} \pm \chi_{_{el}}\, ,\\
    \mathrm{where}\ \chi_{_{el}} &= \arcsin\big( \sin(l)\sin(\delta)+\cos(l)\cos(\delta)\cos(\Theta) \big)\, .
\end{aligned}
\end{equation}
In terms of Jones calculus, the \emph{P-Jones} matrix for parallactic angle rotation for circular polarization is given by \citep[e.g.][]{hales2017}
\begin{equation}
\label{eq:parangmatrix}
    P_p = \begin{pmatrix} \mathrm{e}^{-j\chi_{_p}} && 0 \\  
    0 && \mathrm{e}^{j\chi_{_p}}
\end{pmatrix}.
\end{equation}
Figure \ref{fig:parangplot} shows the evolution of the parallactic angle corresponding to the EHT stations listed in Table \ref{tab:sec3stationlist} over the course of 12 hours towards the direction of M87.
The parallactic angle values of the short baselines formed by (i) JCMT-SMA and (ii) ALMA-APEX evolve similarly due to the proximity of these stations to each other.

\subsubsection{Polarization leakage}
\label{subsubsec:polleak}
The two receptors in the feed are designed to be sensitive to orthogonal polarization states. Ideally, the \emph{D-Jones} terms (or the \emph{D-terms}) corresponding to polarization leakage is represented by a unit matrix, expressed in the appropriate coordinate system (i.e. that of the two polarization states measured by the feed receptors) \citep{hbs1996}. In practice, the non-diagonal terms of this matrix are non-zero, due to the fact that each receptor is sensitive to the opposite polarization state due to electronic or mechanical imperfections in the feed. Hence, the \emph{feed matrix} is given by
\begin{equation}
    \label{eq:feedmatrix}
    \begin{pmatrix} D'_{pR} && d'_{pR} \\  
    -d'_{pL} && D'_{pL}
\end{pmatrix}\, ,
\end{equation}
where $R$ and $L$ denote the two feed receptors (here, in the circular frame). This feed matrix may be decomposed as
\begin{equation}
    \label{eq:gd}
    \begin{aligned}
    \begin{pmatrix} D'_{pR} && 0 \\  
    0 && D'_{pL}
    \end{pmatrix}
    \begin{pmatrix} 1 && d_{pR} \\  
    -d_{pL} && 1
    \end{pmatrix} &= G'_pD_p\, , \\
    \mathrm{where}\ d_{pR} = d'_{pR}/D'_{pR}\, . 
    \end{aligned}
\end{equation}
The matrix $G'_p$ may be absorbed in the \emph{G-Jones} terms that represent the receiver gains (see section \ref{subsec:gandbjones}). The matrix $D_p$ is the feed error or the leakage \emph{D-Jones} term, with the complex numbers $d_{pR}$ and $d_{pL}$ representing the fractional leakage from either feed.

\subsubsection{Implementation}
\label{sss:panddimpl}
\meqsvtwo{} can introduce station-based, frequency-dependent, complex-valued instrumental polarization. The per station complex D-term values for the two polarization feeds are sampled independently for each frequency channel from the normal distribution $\mathcal{N}(d_{p\mu}, d_{p\sigma})$; $d_{p\mu}$ denotes the characteristic empirical leakage value and $d_{p\sigma}$ denotes the scatter for station $p$. The visibilities may be written to the MS in either the sky or the antenna coordinate system (i.e. without or with parallactic angle rotation correction respectively). The visibilities in the two coordinate systems are related by
\begin{equation}
    \label{eq:skyant}
    \mathbfss{V}_{\rm ant} = P_p \mathbfss{V}_{\rm sky} P_q^H\, ,
\end{equation}
where $P_p$ are given by equation (\ref{eq:parangmatrix}). This rotation also includes a constant offset in the feed angle per station. Once this rotation has been applied, the D-terms can be applied to the visibilities in the antenna frame,
\begin{equation}
\label{eq:antframe}
    \mathbfss{V} = D_p (P_p) \mathbfss{V}_{\rm sky} (P_q^H) D_q^H\, .
\end{equation}

If the visibilities are required to be written in the sky frame, then $\mathbfss{V}_{\mathrm{sky}}$ is converted to $\mathbfss{V}_{\mathrm{ant}}$ so that the D-terms may be applied, before being converted back to the sky frame,
\begin{equation}
\label{eq:skyframe}
    \mathbfss{V} = (P_p^H) D_p (P_p) \mathbfss{V}_{\rm sky} (P_q^H) D_q^H (P_q)\, .
\end{equation}
The product $P_p^H D_p P_p$ evaluates to a rotation of the D-terms in the antenna frame by twice the feed angle:
\begin{equation}
\label{eq:pdp}
    P_p^H D_p P_p = \begin{pmatrix} 1 & \exp(2j\chi_p)d_p^R \\ \exp(-2j\chi_p)d_p^L & 1\end{pmatrix}\,.
\end{equation}
The visibilities are usually generated in this frame to correspond to the EHT data.

\subsection{Temporal and spectral variability in receiver gains}
\label{subsec:gandbjones}
The antenna-based gain terms are represented by the complex-valued \emph{G-Jones} matrices which are a generalization of simple antenna gains to polarimetry \citep[e.g.][]{oms2011}. The electronic gains of a pair of circular receptors are given by a diagonal G-Jones matrix in the circular coordinate frame
\begin{equation}
    \label{eq:gjones}
    G_p(t) = \begin{pmatrix} g_{pR}(t) && 0 \\  
    0 && g_{pL}(t)
    \end{pmatrix}\,.   
\end{equation}
Different antenna-based factors such as the $G'$ term in equation (\ref{eq:gd}) may be subsumed into the G-Jones terms. The time-variable complex gains are generated by sampling from the normal distribution $\mathcal{N}(g_{p\mu},\, g_{p\sigma})$ per timestamp, for both the polarization feeds for each station $p$, where $g_{p\mu}$ denotes the characteristic station-dependent gain and $g_{p\sigma}$ denotes the scatter.

Various components along the signal path and the bandpass filters used at each station determine the frequency response of the receiving channels over the observing bandwidth \citep{Syn1999}. This response is not constant over the entire bandwidth and usually falls to about half of the maximum towards both edges of the passband. This results in a frequency-dependent component in the antenna gains which is captured using complex-valued \emph{B-Jones} matrices that vary as a function of frequency. This effect is usually corrected for by observing a bandpass calibrator source whose behaviour across the observing bandwidth is well-known. 

As with the G-Jones terms, the B-Jones matrices for circular receptors are diagonal in a circular coordinate frame,
\begin{equation}
    \label{eq:bjones}
    B_p(\nu) = \begin{pmatrix} b_{pR}(\nu) && 0 \\  
    0 && b_{pL}(\nu)
    \end{pmatrix}\,.
\end{equation}
For each station, \meqsvtwo{} accepts nominal gain values at a few representative frequencies within the bandwidth of the observation, and performs spline interpolation of the bandpass amplitudes for each frequency channel. To these amplitudes, random phases are generated and added to produce the complex quantities. The G-Jones and B-Jones terms are simulated separately to provide independent, fine-grained variability of gains along the time and frequency axes.

\subsection{Noise considerations}
\label{subsec:thermalnoise}
\subsubsection{Thermal noise}
Noise due to various factors such as receiver electronics, atmospheric emission, background radiation and ground (i.e. spillover) radiation affect the system sensitivity adversely. The system temperature, $T_{{\mathrm sys}}$, equivalent to the power per unit frequency due to the noise, $P_N$, accounts for this noise and is given by
\begin{equation}
\label{eq:tsys}
T_{{\mathrm sys}} = \frac{P_N}{k_B},
\end{equation}
where $k_B$ is the Boltzmann constant\footnote{The tropospheric contribution to the increase in $T_{\mathrm{sys}}$ (section \ref{sss:meantrop}) is added to the noise budget as mentioned below.}. The system temperature is often expressed in terms of the system equivalent flux density (SEFD), which is defined as the flux density of a source that would deliver the same amount of power as the system
\begin{equation}
    \mathrm{SEFD} = \frac{2 k_{\mathrm B} T_{\mathrm{sys}}}{\eta A_{\mathrm e}}\, ,
\end{equation}
where $\eta$ is the antenna efficiency and $A_{\mathrm e}$ is the effective area of the antenna. \meqsvtwo{} accepts the station-dependent SEFD values as inputs from which the per-baseline rms uncertainty, $\sigma_{pq}$,  on a visibility in units of Jy is computed \citep{TMS2017}:
\begin{equation}
\label{eq:visnoise}
\sigma_{pq} = \frac{1}{\eta} \sqrt{\frac{{\mathrm{SEFD}}_p {\mathrm{SEFD}}_q}{2\Delta \nu \tau}}\, ,
\end{equation}
where $A_{\mathrm e}$ denotes the effective area of the telescope and $\eta$ comprises
any relevant efficiency terms, such as the antenna aperture efficiency, $\eta_{\rm ap}$, the correlator efficiency, $\eta_{\rm corr}$, $\Delta\nu$ is the bandwidth and $\tau$ the integration time. For standard 2-bit quantization, $\eta$ is set to 0.88. Since the system noise is broadband and mostly stationary, it can be described using a Gaussian distribution and the uncertainty in their measurements can be reduced by increasing the number of independent measurements $N = 2\Delta \nu \tau$.

In the RIME implementation, this thermal noise becomes an additive term per polarization, distributed normally with zero mean and a variance of $\sigma_{pq}^2$ per visibility:
\begin{equation}
\label{eq:noiserime}
\mathbfss{V}_{pq} = \mathbfss{V}'_{pq} + \mathcal{N}(0, \sigma_{pq}^2).
\end{equation}
This additive thermal noise matrix has the same dimensionality as the data, varying with time, baseline, frequency, and polarization.

\subsubsection{Visibility weighting}
\label{ss:viswt}
The MS format allows for the estimated rms noise values and visibility weights to be recorded alongside the data. The $\sigma_{pq}$ values computed above are used to generate per-visibility baseline-dependent thermal noise. These terms are added with the increase in the sky brightness temperature due to the troposphere (section \ref{sss:meantrop}) and are used to fill the SIGMA and SIGMA\_SPECTRUM columns in the MS. Inverse-variance weighting is used to fill in the visibility weights columns WEIGHT and WEIGHT\_SPECTRUM.

\section{PolSolve: polarization Leakage Estimation}
\label{sec:polsolve}
We use \polsol{} to validate the instrumental polarization capabilities of \meqsvtwo{}. \polsol{} is part of \poltools{}\footnote{\url{https://launchpad.com/casa-poltools}.}, the polarimetry toolbox developed for \casa{}, aimed at the simulation, calibration, and basic analysis of polarimetric VLBI observations \citep{polsolveRef}. It uses the full RIME, simplified for the case of narrow-field observations, to estimate and correct for instrumental polarization using observations of spatially-resolved polarization calibrators.

\polsol{} has many advantages compared to the \aips{}\footnote{\url{http://www.aips.nrao.edu}.} task \lpcal{} \citep[][]{leppanen1995}, the main algorithm used by the VLBI community for polarimetric calibration. It uses a non-linear model of polarization leakage derived from the full RIME for handling high leakage values. \polsol{} can also model the frequency-dependent variations in D-terms to enable calibration of wide fractional bandwidths and perform D-term estimates based on cross-polarization self-calibration. We take advantage of many of these features in performing the validation tests for this paper.

The \aips{}-based \textsc{gpcal}\footnote{\url{https://github.com/jhparkastro/gpcal}.} pipeline also addresses many of the shortcomings of \lpcal{} \citep{park2021} and has been shown to be consistent with \polsol{} \citep{eht7}. A detailed discussion of the impact of these features on the quality of VLBI polarimetry are discussed in \citet{polsolveRef}. Below, we give a brief description of the procedure and equations used by \polsol{}.

\subsection{The POLSOLVE fitting parameters}
\label{subsec:polsolpars}
\polsol{} uses the Levenberg-Marquardt algorithm \citep{numrecipes2007} to minimize the error function $\chi^2(\boldsymbol{x})$, where $\boldsymbol{x}$ consists of parameters that model both source and instrumental polarization, divided into two subsets. It uses two different approaches to account for the polarization structure of a source, both of which assume that the source brightness distribution can be described as a linear combination of source components \citep[e.g., point sources, for the case of \clean{} deconvolution,][]{hogbom1974}.

The first approach divides the set of CLEAN components $\{I_i\}$ into $N_s$ subsets, also called source ``subregions'', for which a constant fractional polarization is assumed. The corresponding model visibility functions for Stokes $Q$ and $U$ are given by
\begin{equation}
\label{eq:visqmodel}
\mathrm{V}_Q = \sum_s^{N_s} \left( q_s \sum_k^{n_s}{\mathcal{F}(I_k^s)} \right) ~~\mathrm{and}~~ \mathrm{V}_U = \sum_s^{N_s} \left( u_s \sum_k^{n_s}{\mathcal{F}(I_k^s)} \right)\,, 
\end{equation}
where $N_s$ is the number of subregions with constant fractional polarization, $n_s$ is the number of CLEAN components inside the $s$-th subregion, and $I_k^s$ is $k$-th CLEAN component of Stokes $I$ in the $s$-th subregion. The vectors $\boldsymbol{q} = \{q_s\}$ and $\boldsymbol{u} = \{u_s\}$ are the parameters that \polsol{} fits for.

The second approach uses independent, and fixed ($N_s=0$), source models for the brightness distributions of $Q$ and $U$. The model visibilities for $Q$ and $U$ then take the simple form
\begin{equation}
\label{eq:visqmodel2}
\mathrm{V}_Q = \sum_i^{N_q}{\mathcal{F}(Q_i)} ~~\mathrm{and}~~\mathrm{V}_U = \sum_i^{N_u}{\mathcal{F}(U_i)}\,,
\end{equation}
where $\{Q_i\}$ and $\{U_i\}$ are the (e.g. \clean{}) components used to model the polarization source structure. In this case, \polsol{} performs a cross-polarization self-calibration estimate of the D-terms.

Instrumental polarization is represented by two complex quantities per antenna, corresponding to each polarization. The leakage terms obey the equation
\begin{equation}
\label{eq:PolSolveEq}
\begin{aligned}
\begin{pmatrix} \mathrm{V}_{RR}^{i} && \mathrm{V}_{RL}^{i} \\ \mathrm{V}_{LR}^{i} && \mathrm{V}_{LL}^{i} \end{pmatrix} =  \begin{pmatrix} 1 && D_R^p \\ D_L^p && 1 \end{pmatrix} \begin{pmatrix} \mathrm{V}_I^{uv} && \mathrm{V}_P^{uv} \\  \mathrm{V}_{P}^{*uv} && \mathrm{V}_I^{uv} \end{pmatrix} \\
\begin{pmatrix} 1 && (D_L^*)^q \\ (D_R^*)^q && 1    \end{pmatrix},
\end{aligned}
\end{equation}
where $\mathrm{V}_{k}^{i}$ is the $i$-th observed visibility of polarization product $k$ (where $k$ is one of $RR$, $RL$, $LR$, or $LL$) and $u$ and $v$ are the coordinates in Fourier space. We construct the functions for the complex polarization vector, $\boldsymbol{P}$, in the form 
\begin{equation}
\label{eq:polvisrep}
\mathrm{V}_P = \mathrm{V}_Q + j\mathrm{V}_U  ~~~ \mathrm{and} ~~~ \mathrm{V}_{P^*} = \mathrm{V}_Q - j\mathrm{V}_U.
\end{equation}
These visibilities are assumed to be fully calibrated for atmospheric effects and electronic antenna gains (computed in the frame of the antenna mounts).


\section{Simulating mm-wave observations}
\label{sec:demo}
\meqsvtwo{} was developed primarily for generating synthetic data for the EHT at mm-wavelengths, but it can equally well be applied to any array, including proposed arrays such as ngEHT and ngVLA. For instance, \meqsvtwo{} can be used to perform a more elaborate exploration of the VLBI capabilities of MeerKAT for performing extragalactic surveys presented in \citet{deane2016}. \meqsvtwo{} has also been used to generate 5 GHz synthetic EVN observations for simulating phase corruptions affecting astrometric uncertainties \citep{vanlangevelde2019}. \citet{roelofs2020} use \meqsvtwo{} for generating uncalibrated synthetic data observed using an extended EHT array with additional stations located at potential future sites.

\begin{figure}
 \includegraphics[scale=0.325]{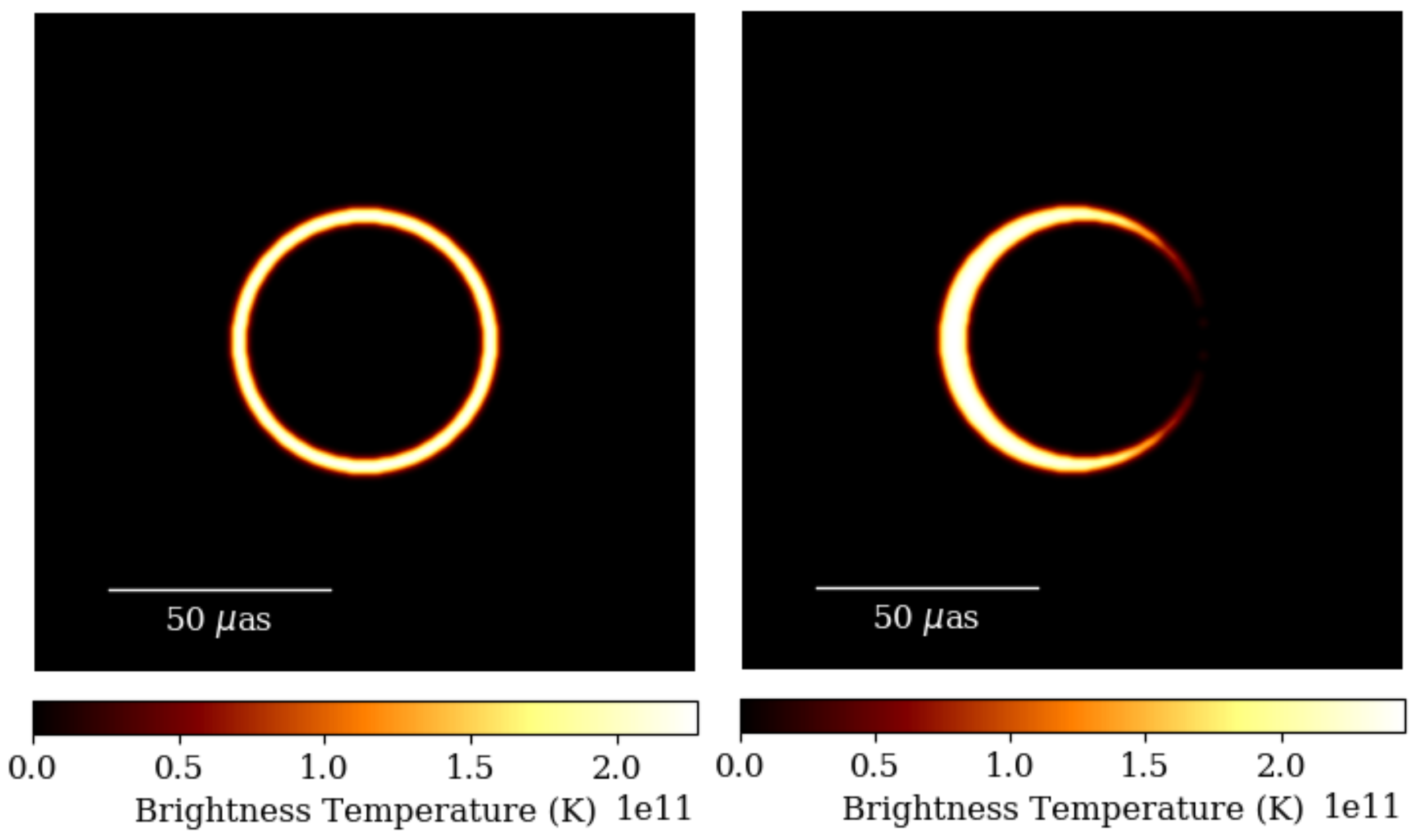}
 \caption[bpskies]{Symmetric ring (\emph{left}) and asymmetric crescent (\emph{right}) models from \citet{kamruddin2013} used in generating synthetic data for sections \ref{sec:demo} and \ref{sec:bandpass}.}
\label{fig:bpskies}
\end{figure}

We generate synthetic data with \meqsvtwo{} for three mm-wavelength arrays: EHT2017 array, ngVLA in its long baseline array configuration, and ALMA array in its extended configuration. The EHT2017 array consists of the stations shown in Table \ref{tab:sec3stationlist} observing at a frequency of 230 GHz. The ngVLA array consists of all 18 stations in the Long Baseline Array (LBA) and the central core reduced to a single high-sensitivity site, observing at 86 GHz. ALMA consists of 43 12-metre diameter antennas in its most extended configuration, observing at 230 GHz. An asymmetric crescent \citep{kamruddin2013} created with \ehtim{}, with the inner radius offset by 3 $\mu$as in the horizontal direction was used as the source model (Figure \ref{fig:bpskies}, \emph{right}).

Various propagation path effects described in the previous section such as pointing offsets, tropospheric effects, receiver gains, polarization leakage, and thermal noise are introduced. Nominal pointing offsets based on a priori station information are used \citep[e.g.][]{roelofs2020}. The aperture efficiencies were determined using targeted observations of planets of known brightness temperatures as calibrators \citep{eht3}. The individual station SEFDs are determined by extrapolating measured system temperatures to zero airmass \citep{janssen2019,roelofs2020}.

We adopt the median values measured by \vlbimon{}\footnote{\url{https://bitbucket.org/vlbi}.} at the individual sites during the 2017 EHT observing campaign \citep{eht2} for the weather parameters. For stations for which this information does not exist, it was obtained through climatological modelling using the data sets from Modern-Era Retrospective Analysis for Research and Applications, version 2 (MERRA-2) from the NASA Goddard Earth Sciences Data and Information Services Center \citep[GES DISC;][]{gelaro2017} and the \textsc{am} atmospheric model software \citep{paine2019}. The atmospheric conditions over all the ALMA antennas are the same, though independently simulated for each antenna.

\begin{figure*}
 \includegraphics[scale=0.27]{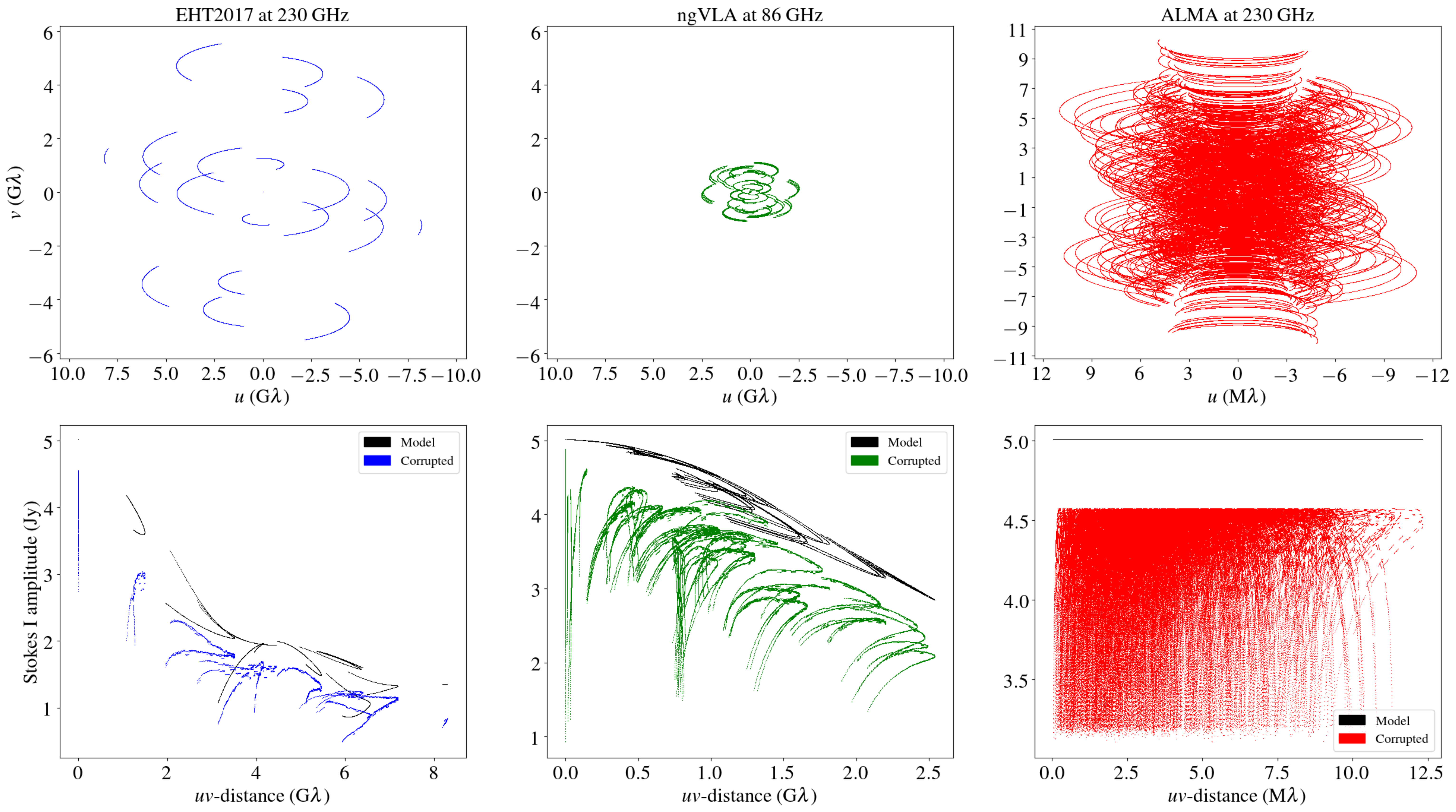}
 \caption[compare]{\emph{Top row}: The \emph{uv}-coverages of EHT2017, ngVLA, and ALMA towards M87; note that the ALMA baselines are three orders of magnitude shorter than the other two. \emph{Bottom row}: Model (black) and corrupted Stokes $I$ visibility amplitudes as a function of \emph{uv}-distance, as observed by each array. ALMA, due to its relatively short baselines, does not resolve the source and the model visibilities correspond to a horizontal line at 5 Jy.}
\label{fig:compare_uvcov_vis}
\end{figure*}

The top row of Figure \ref{fig:compare_uvcov_vis} shows the \emph{uv}-coverages for the three arrays in the direction of M87 at their respective observing frequencies, showing how the different arrays complement each other.
The bottom row of Figure \ref{fig:compare_uvcov_vis} shows both the corrupted and uncorrupted Stokes $I$ visibility amplitudes for each array as a function of \emph{uv}-distance. The 60 $\mu$as crescent is fully resolved by EHT2017 as expected. At 86 GHz, ngVLA resolves it partially, clearly seeing the source as having some extended structure. ALMA, with its longest baselines of only ~12 km, sees the crescent as a point source at 230 GHz and the model visibilities show up as a straight line at 5 Jy in the figure. In the following sections, we concentrate only on the EHT2017 array.

\section{Effect of bandpass on closure quantities}
\label{sec:bandpass}
\begin{figure}
 \includegraphics[scale=0.525]{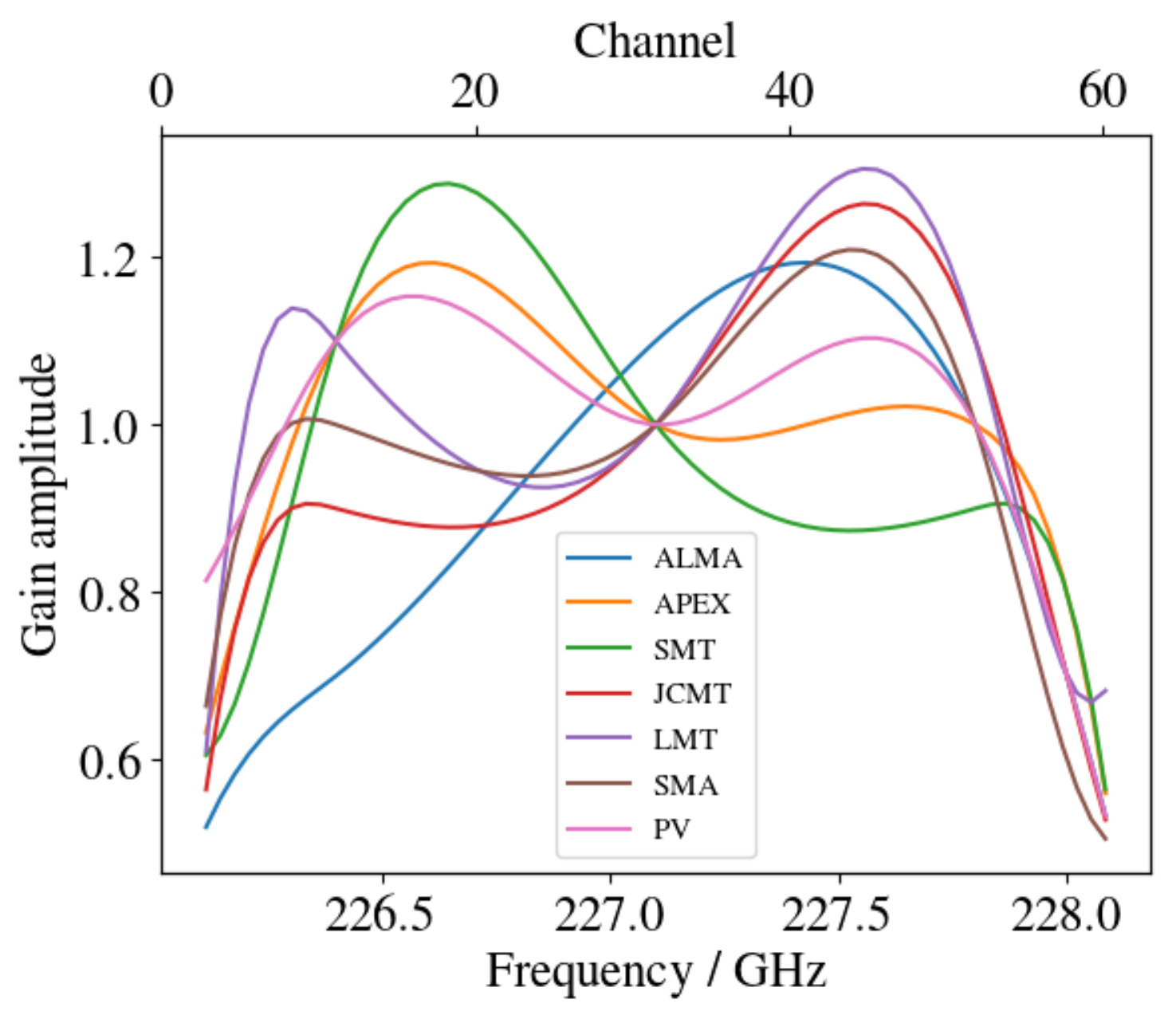}
 \caption[bpshapes]{Station-based bandpass amplitudes used in generating synthetic data for section \ref{sec:bandpass}.}
\label{fig:bpshapes}
\end{figure}

In this section, we estimate the magnitude of systematic non-closing errors introduced by complex bandpass gains in EHT observations at 230 GHz using synthetic data. We introduce station-based bandpasses of varying shapes, spline-interpolated to all frequency channels as shown in Figure \ref{fig:bpshapes}.
The phases are sampled conservatively within the uniform range of $\pm 30^{\circ}$ (Michael Janssen, private comm). Two different sky models from \citep{kamruddin2013}, generated using \ehtim{} were used: a symmetric ring with outer radius 30 $\mu$as and width 3 $\mu$as, and an asymmetric crescent with the inner radius offset by 3 $\mu$as in the horizontal direction (Figure \ref{fig:bpskies}).
The total flux of the ring was set to 5 Jy to ensure that the S/N of the data was high so that systematic non-closing errors could be unambiguously detected.
Thermal noise generated using equation (\ref{eq:visnoise}) was applied to all datasets. All synthetic data sets used in this subsection are generated with a time resolution of 1 s and a frequency resolution of 31.25 MHz per frequency channel, with 64 channels, spanning a bandwidth of 2 GHz. These data are then band-averaged to a single channel of width 2 GHz. We generated four categories of synthetic data, with 10 data sets to each category, differing in the realisation of thermal noise:
\begin{enumerate}
    \item symmetric ring with only thermal noise corruption,
    \item symmetric ring with thermal noise and bandpass corruption,
    \item asymmetric crescent with only thermal noise corruption, and
    \item asymmetric crescent with thermal noise and bandpass corruption.
\end{enumerate}

Closure quantities are formed around a closed loop of stations to eliminate the effects of station-based gain errors \citep[e.g.][]{Blackburn2020}.  Closure phases are formed over a closed triangle $pqr$ as
\begin{equation}
\label{eq:cp}
    \psi_{C,pqr} = \mathrm{arg}\ (\mathrm{V}_{pq}\mathrm{V}_{qr}\mathrm{V}_{rp}),
\end{equation}
and closure amplitudes are formed over a quadrangle of stations $pqrs$:
\begin{equation}
    \label{eq:logca}
    \mathrm{ln}\ A_{C,pqrs} = \mathrm{ln}\ \frac{\mathrm{V}_{pq}\mathrm{V}_{rs}}{\mathrm{V}_{pr}\mathrm{V}_{qs}},
\end{equation}
where ``ln'' is the natural logarithm. The closure phases (CP) and the log closure amplitudes (LCA) are susceptible to systematic non-closing errors. The intra-site baselines ALMA-APEX and JCMT-SMA enable one to form ``trivial'' closure quantities, between different combinations of these four stations. The trivial CP and LCA so formed should ideally be zero, but factors such as band-averaging without correcting for bandpass errors and intrinsic large-scale asymmetry in source structure can break the assumptions of trivial closure.

We evaluate the magnitude of systematic non-closing errors in trivial closure quantities following the EHT data validation procedures outlined in \citet{eht3} and in \citet{wielgus2019}. We employ the $\mathrm{MAD}_0$ (median absolute deviation from zero) statistic, which, for a normally distributed variable $Y$ with zero mean takes the form
\begin{equation}
\label{eq:mad0}
    \mathrm{MAD}_0(Y) = 1.4826\ \mathrm{median}(|Y|),
\end{equation}
where the subscript 0 denotes that the raw moment is estimated. The scaling factor 1.4826 ensures that $\mathrm{MAD}_0$ acts as an unbiased estimator of standard deviation for normally distributed data. The total uncertainty, $\sigma$, associated with the trivial closure quantities defined above is given by
\begin{equation}
    \label{eq:mad0sigma}
    \sigma^2 = \sigma^2_{\mathrm{th}}\ +\ s^2,
\end{equation}
where $\sigma_{\mathrm{th}}$ is the known thermal component 
and $s$ denotes the systematic non-closing error, modeled as a constant. For a trivial closure quantity $X$, we solve for the characteristic value of $s$ by enforcing a following condition:
\begin{equation}
    \label{eq:mad0x}
    \mathrm{MAD}_0\Bigg(\frac{X}{\sigma}\Bigg) = \mathrm{MAD}_0\Bigg(\frac{X}{\sqrt{\sigma^2_{\mathrm{th}}\ +\ s^2}}\Bigg) = 1.
\end{equation}

The $\mathrm{MAD}_0$ values estimated using equation (\ref{eq:mad0}) and the characteristic magnitude of systematic errors $s$ calculated with equation (\ref{eq:mad0x}) for the four data sets are given in Table \ref{tab:bandpasses}.
\begin{table*}
\centering
\caption{Estimated $\mathrm{MAD}_0$ values and characteristic magnitudes of systematic errors in the trivial closure quantities in the synthetic data sets described in section \ref{sec:bandpass}.}
\label{tab:bandpasses}
\begin{tabular}{lcccc}
\hline
 & \multicolumn{2}{c}{\rule{0pt}{1em}Trivial CP} & \multicolumn{2}{c}{Trivial LCA} \\ \cline{2-5}
\rule{0pt}{1em}Type of data set & $\mathrm{MAD}_0$ & $s$ (deg.) & $\mathrm{MAD}_0$ & $s$ (\%) \\ \hline
\rule{0pt}{1em}(i) Symmetric ring (thermal noise only) & $1.012\pm0.042$ & $0.041\pm0.011$ & $1.148\pm0.051$ & $0.4\pm0.1$ \\
\rule{0pt}{0.6em}(ii) Symmetric ring (thermal noise + bandpass gains) & $2.76\pm0.408$ & $0.681\pm0.088$ & $4.074\pm0.361$ & $4.7\pm0.5$ \\
\rule{0pt}{0.6em}(iii) Asymmetric crescent (thermal noise only) & $1.118\pm0.067$ & $0.077\pm0.016$ & $0.984\pm0.04$ &
$0.08\pm0.02$ \\
\rule{0pt}{0.6em}(iv) Asymmetric crescent (thermal noise + bandpass gains) & $3.862\pm0.363$ & $0.684\pm0.071$ & $7.484\pm0.829$ & $3.8\pm0.5$ \\
\hline
\end{tabular}
\end{table*}
 The uncertainties are estimated by obtaining these values for 10 data sets with different thermal noise realisations.
 
If the error budget is exactly accounted for, then the $\mathrm{MAD}_0$ estimator is equal to 1. This is evident from the estimated $\mathrm{MAD}_0$ values being very close to unity when thermal noise is the only corruption introduced. The corresponding systematic errors are about $0.08^{\circ}$ for trivial closure phases and less than $0.4$ per cent for trivial log closure amplitudes. In the presence of bandpass gains that are not accounted for in the error budget, the $\mathrm{MAD}_0$ values and the systematic errors increase noticeably. For the trivial closure phases, the $\mathrm{MAD}_0$ values indicate that the reported errors are too small by factors of 2.76 and 3.86, for the symmetric ring and asymmetric crescent models respectively; for trivial log closure amplitudes, these factors increase to about 4 and 7.5 respectively. The estimated systematics are about $0.7^{\circ}$ for trivial closure phases and up to $5$ per cent for trivial log closure amplitudes. 

For the scale of bandpass gains assumed here, these errors are comparable to the magnitude of the systematic non-closing errors for EHT 2017 observations of 3C279 and M87 ($2^{\circ}$ for closure phases and less than $4\%$ for log closure amplitudes) quantified by \citet{eht3}. Alongside the many factors such as intrinsic source structure or polarisation leakage, this indicates how unaccounted-for bandpass errors can have a similar effect on the closure quantities. The spectral capabilities of \meqsvtwo{} will enable us to perform a thorough study of EHT bandpass characteristics and their effect on the observed data.


\section{Simulating instrumental polarization}
\label{sec:poldemo}

\subsection{Simulating polarized sources with leakage}
\label{ss:estimateleakage}
\begin{figure*}
    \centering
    \includegraphics[scale=0.34]{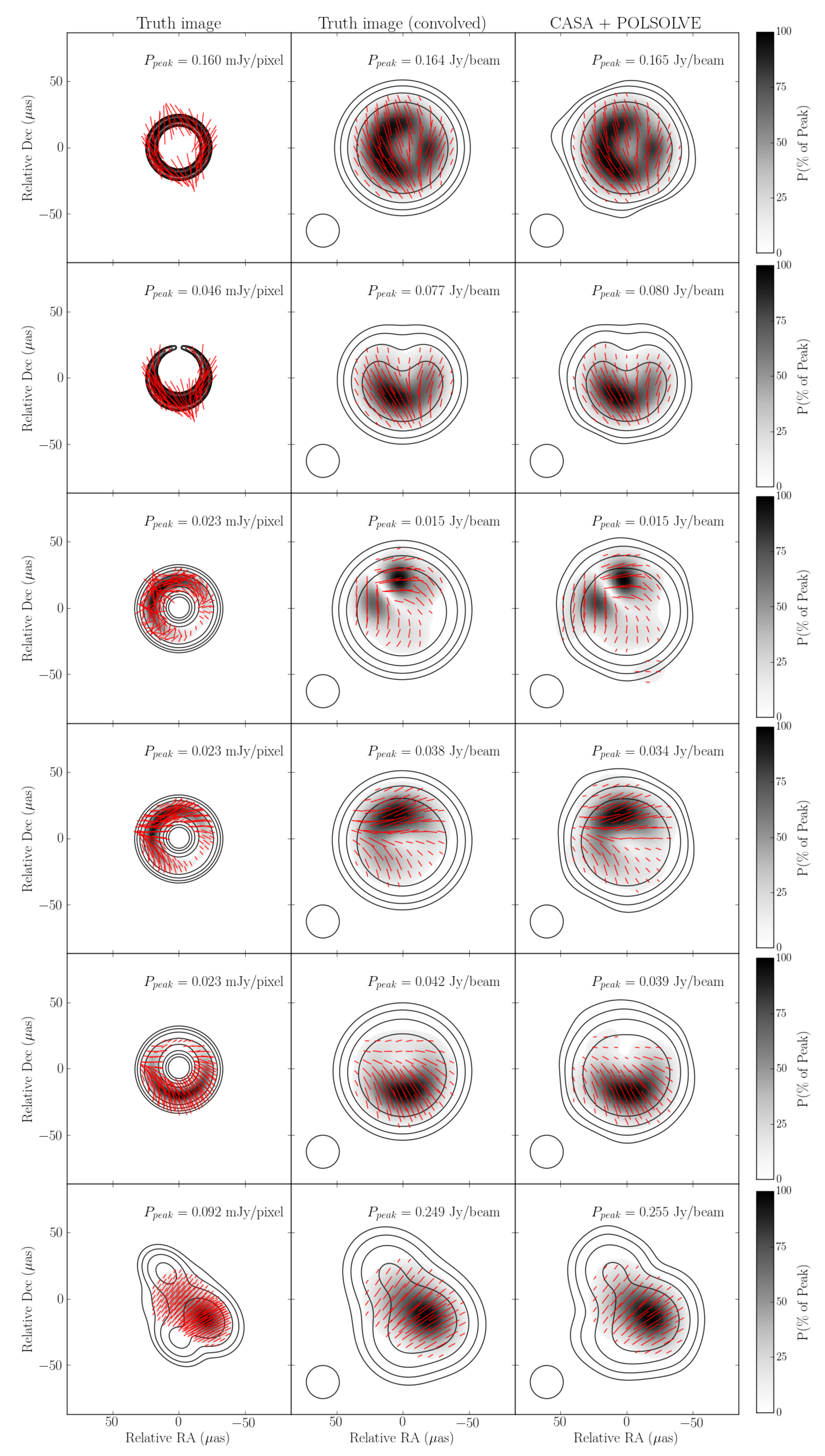}
    \caption{From left to right: source model used in the simulations, model convolved with the nominal EHT beam of FWHM 25\,$\mu$as, and CLEAN model obtained from the synthetic data convolved with the same 25\,$\mu$as beam. Contours indicate total intensity and are logarithmically spaced between 1 and 100 per cent of the intensity peak; grayscale images show polarization intensity; lines indicate the EVPA distribution, with lengths proportional to the local fractional polarization.}
    \label{fig:polsolveimages}
\end{figure*}

To validate the implementation of instrumental polarization in \meqsvtwo{} using \polsol{}, we generate six synthetic data sets, corrupted with instrumental polarization, time-varying complex gains, and thermal noise.
The accuracy to which D-terms can be estimated depends on various factors such as station parameters and intrinsic source polarization structure. The source models we use are shown in the first column of Figure \ref{fig:polsolveimages}. The truth images convolved with the nominal EHT beam are shown in the second column. Source model 1 is a highly polarized ring while model 2 is an offset crescent with low polarization. In both models, the polarized structure closely follows the Stokes I distribution. Models 3-5 are ring sources with differing EVPA structures. Model 6 represents the characteristic source structure of any VLBI calibrator source at cm and mm-wavelengths. We choose models differing in EVPA structures and fractional polarization. All source model images are generated using \ehtim{}.

The polarization calibration is performed by running \polsol{} in iterative self-calibration mode. The D-terms for ALMA-APEX and, optionally, JCMT-SMA, are solved for using a point source model and are fixed for the rest of the calibration process. A model of the source is generated using \clean{} and multiple iterations of phase and amplitude self-calibration are performed with decreasing solution intervals. Finally, several iterations of \polsol{} and \clean{} are performed, by fixing the full polarization model to the output of \clean{} at each iteration.

\begin{figure*}
 \includegraphics[scale=0.5]{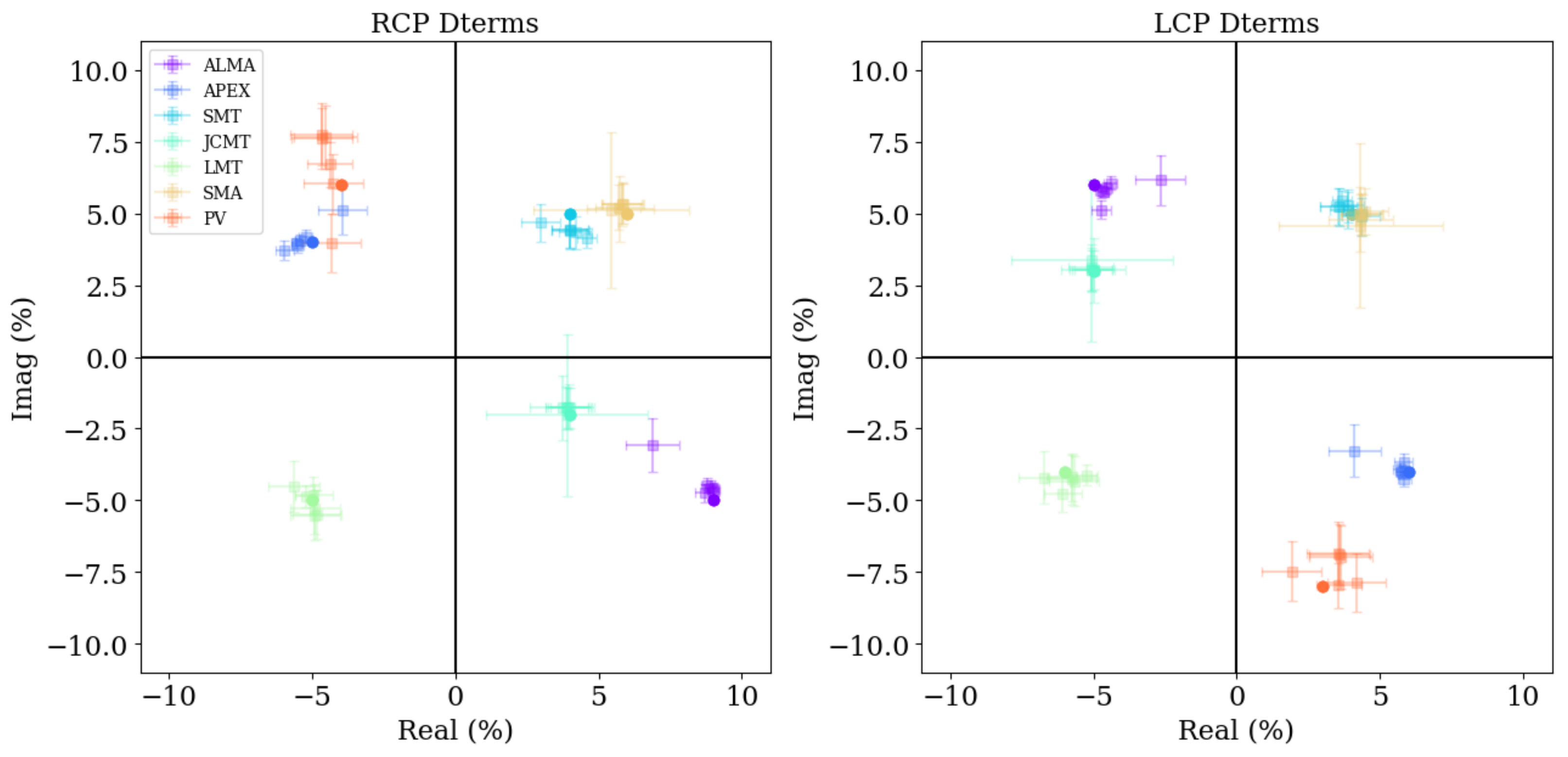}
 \caption[dterms]{D-term estimates for all stations from \polsol{} in iterative self-calibration mode for synthetic data sets described in section \ref{ss:estimateleakage}. Each square with error bar corresponds to one data set. The filled circles denote the ground-truth D-term values.}
\label{fig:dterms2}
\end{figure*}

The reconstructed images convolved with the \clean{} beam are shown in the third column of Figure \ref{fig:polsolveimages}.
The polarization peaks displayed in each panel are found to correlate well between the ground-truth images and the reconstructed images, as are the Stokes I distribution and the polarization structure. Figure \ref{fig:dterms2} shows the D-terms recovered by \polsol{} for each data set for all stations, along with the ground-truth D-term values.
The recovered D-terms correlate highly with the ground-truth values. The recovered D-terms for model 1 are the least accurate owing to its high intrinsic fractional polarization. D-terms of similar magnitude applied to models with low fractional polarization are estimated more accurately. The D-term estimates for the stations participating in intra-baseline fitting (ALMA-APEX and JCMT-SMA) are the most tightly constrained and those corresponding to the station with the longest baselines on a short ($u, v$) track, PV, have the largest dispersion.

\subsection{Frequency-dependent polarization leakage}
\label{ss:freqdterms}
\meqsvtwo{} can also introduce frequency-dependent variations in the simulated D-terms. The D-terms were generated independently for each channel by sampling them from a normal distribution. We use \polsol{} in iterative self-calibration mode to solve for these D-terms, treating them as independent complex numbers per frequency channel (see section \ref{sec:polsolve}). 

\begin{figure}
 \includegraphics[scale=0.39]{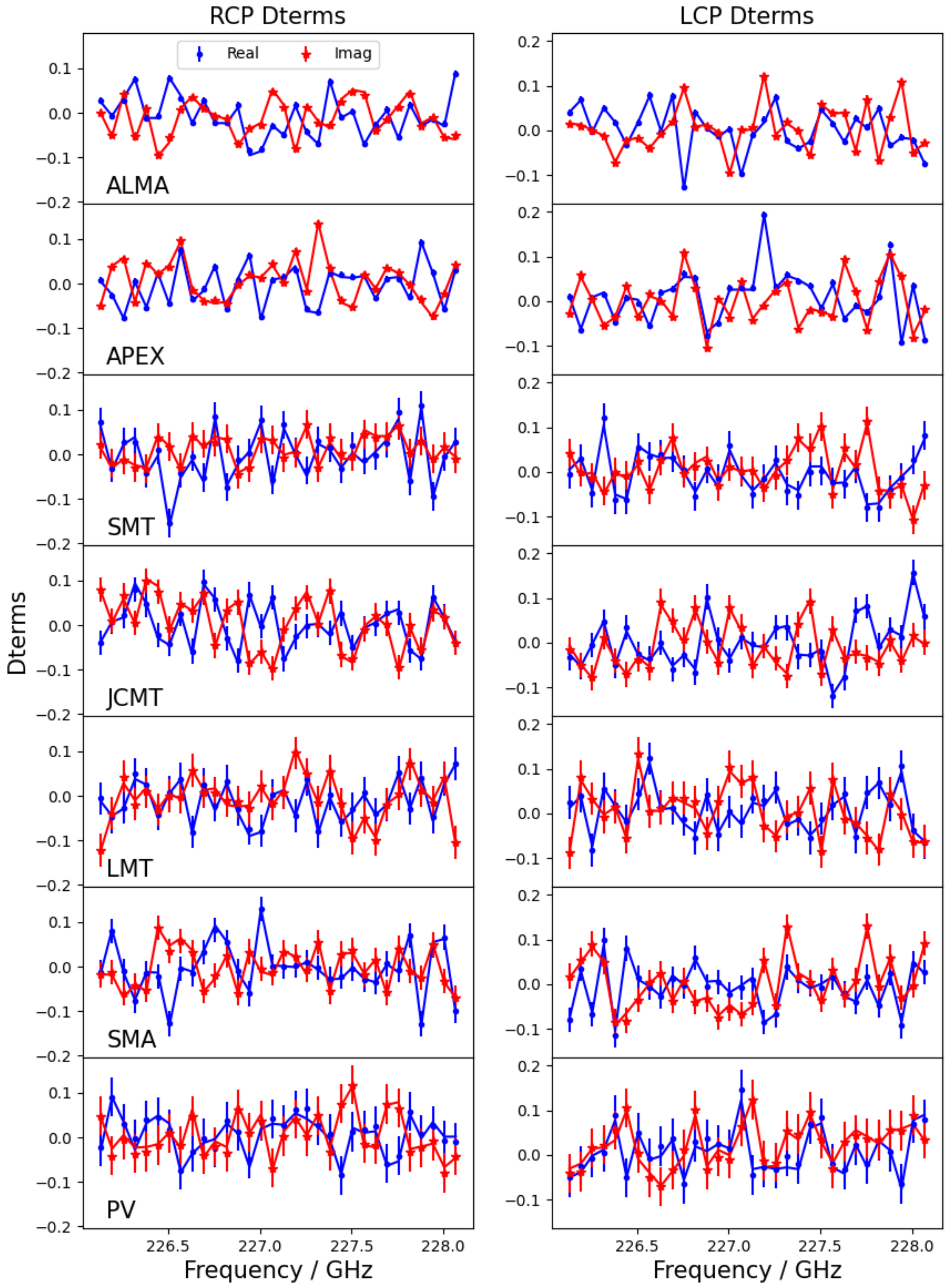}
 \caption[dterms]{\polsol{} estimates of per-channel D-terms in polarization self-calibration mode. The solid lines represent the ground-truth values, while the circles and asterisks represent the \polsol{} estimates.}
\label{fig:freqdterms}
\end{figure}

We use Model 3 from Figure \ref{fig:polsolveimages} to generate synthetic data with 2 GHz bandwidth divided into 32 spectral windows, with frequency-varying D-terms introduced alongside complex gains and thermal noise. Figure \ref{fig:freqdterms} shows the simulated D-terms and the D-terms recovered by \polsol{} with error bars.
The estimated D-terms correlate remarkably well with the simulated values. The magnitude of the errors are similar to that corresponding to model 3 in Figure \ref{fig:dterms2}.

\section{Discussion}
\label{sec:disc}
Synthetic observations provide uniform data sets with known source and instrumental properties for performing various kinds of feasibility studies. Since \meqsvtwo{} aims to compute data corruptions from first principles, it can be used to perform a systematic exploration of realistic site and antenna parameters, which helps in commissioning new sites and optimizing observing schedules \citep[e.g.][]{Blackburn2020}. Conversely, with known site and antenna characteristics, the synthetic data can be used to quantitatively compare different astrophysical source models that characterize the observed emission.

The realistic synthetic data provided by \meqsvtwo{} can be used to validate new imaging, calibration, and parameter estimation techniques. Synthetic data generated with \meqsvtwo{} have been used to test \textsc{zagros}, a fully-Bayesian fringe-fitting framework for analysing VLBI observations \citep{iniyan2020}. These data can potentially be used to compare different parameter estimation frameworks currently being used for analysing EHT data, such as \textsc{themis} \citep{broderick2020}, \textsc{march} \citep{psaltis2020}, and \textsc{dmc}\footnote{\url{https://github.com/dpesce/eht-dmc}.} \citep{pesce2021}. Fully end-to-end pipelines starting from synthetic data generation with \meqsvtwo{}, followed by a priori calibration and posterior estimation using statistical visibility analysis packages are under active development to enable large-scale feasibility studies. The modular nature of the software lends itself easily to constructing such end-to-end pipelines that allow calibration to start further up the signal chain \citep[e.g.][]{iniyan2020}. The computational requirements for such large-scale studies necessitate adapting these software packages to a High Performance Computing (HPC) environment.

The experiments performed in this paper are intended to illustrate the capabilities of \meqsvtwo{}. More elaborate studies of the various propagation path effects on calibration and imaging must be performed for future site selection and testing the capabilities and limitations of new algorithms.
More exhaustive exploration of frequency-dependent gain variations and their effects on closure quantities (see section \ref{sec:bandpass}) need to be performed to fully understand the bandpass characteristics of the EHT array and solve for them.

Section \ref{sec:poldemo} illustrates how fractional polarization intrinsic to the source and the lack of short baselines to a station can affect the recovery of station-based instrumental polarization. The accuracy of the measured D-terms can be improved by performing a self-consistent multi-source fit to the D-terms, if observations of multiple sources are available for the same epoch \citep[e.g.][]{polsolveRef}. More complex experiments devised to study this effect will help to fully exploit the full-polarization data products generated by EHT observations.

\section{Summary and Outlook}
\label{sec:final}
We present \meqsvtwo{}, a synthetic data generation software package capable of performing synthetic VLBI observations with full polarimetric, time and frequency-resolved astrophysical models and realistic propagation path effects. It can also be seamlessly integrated with calibration tools such as the EHT \textsc{casa} pipeline to perform additional calibration steps to mimic VLBI data at different stages along the propagation path.

We conduct preliminary investigations of how source model asymmetry affects closure quantities in the presence of complex bandpass effects. We also validate the polarimetric capabilities of \meqsvtwo{} by generating synthetic EHT observations of polarized geometric source models with instrumental polarization and estimating the D-terms using \polsol{}, showing that the reconstructed images correlate very well with input source models.

Future versions will be able to simulate more polarimetric effects such as Faraday rotation in the ISM and ionospheric phase corruption effects for simulating cm-VLBI radio observations. The ability to input more time and frequency-variable weather parameters will also be introduced. We also plan to take full advantage of new distributed and parallel computing algorithms and software packages for generating synthetic data sets on a large scale much faster.

\meqsvtwo{} currently optionally splits large data sets into subsets for processing, to accommodate systems with low memory specifications. Such features can easily be ported to forward-modelling software packages such as \codex{} (Perkins et al. in prep) which provide distributed CPU and GPU computing functionality.
For handling large data sets, we plan to use \daskms{} (Perkins et al. in prep), which uses \dask{}, an open source library for parallel programming in Python \citep{rocklin2015dask}, to scale computations.
\daskms{} can convert MS v2.0 data between the \casa{} table format, and other high performance, cloud-native formats such as \textsc{parquet} \citep{vohra2016parquet} and \textsc{zarr} \citep{miles2021zarr}. These new formats are explicitly designed for parallel, distributed processing and offer superior disk I/O performance, which will accelerate the synthetic data generation process. Finally, we also plan to set up a publicly available online interface to \meqsvtwo{}, which can be of use to students and teachers alike without them having to invest in the requisite computing power.

\section*{Acknowledgements}
We thank Benna Hugo and Andr\'e Offringa for discussions on \textsc{wsclean} and Mareki Honma for his valuable comments. IN and RPD acknowledge support from the New Scientific Frontiers with Precision Radio Interferometry group grant awarded by the South African Radio Astronomy Observatory (SARAO), which is a facility of the National Research Foundation (NRF), an agency of the Department of Science and Innovation (DSI). IN, RPD, and OS acknowledge funding by the South African Research Chairs Initiative of the DSI/NRF. IMV acknowledges support from Research Project PID2019-108995GB-C22 of Ministerio de Ciencia e Innovaci\'on (Spain) and from the GenT Project CIDEGENT/2018/021
of Generalitat Valenciana (Spain). FR, MJ, JD, MM, and HF acknowledge support from the ERC Synergy Grant ``BlackHoleCam: Imaging the Event Horizon of Black Holes" (Grant 610058). FO and DP acknowledge support from the NSF PIRE grant OISE-1743747. MW and LB were supported by the Black Hole Initiative at Harvard University, which is funded by grants from the John Templeton Foundation and the Gordon and Betty Moore Foundation to Harvard University. KLB acknowledges support from the MSIP2: NSF Award 2034306. GB acknowledges support from the Ministero degli Affari Esteri della Cooperazione Internazionale - Direzione Generale per la  Promozione del Sistema Paese Progetto di Grande Rilevanza ZA18GR02 and the National Research Foundation of South Africa (Grant  Number 113121) as part of the ISARP RADIOSKY2020 Joint Research Scheme.

\section*{Data Availability}
\meqsvtwo{} is open source and publicly available on GitHub at \url{https://github.com/rdeane/MeqSilhouette}. Its documentation can be found at \url{https://meqsilhouette.readthedocs.io}. Bugs and other issues can be reported by creating an issue on the repository.

The data underlying this article will be shared on reasonable request to the corresponding author.



\bibliographystyle{mnras}
\bibliography{references}






\bsp	
\label{lastpage}
\end{document}